# Aerodynamics of Rotor Blades for Quadrotors

Moses Bangura*, Marco Melega**, Roberto Naldi** and Robert Mahony*

*Australian National University, **University of Bologna

*{Moses.Bangura, Robert.Mahony}@anu.edu.au, **{marco.melega3,roberto.naldi}@unibo.it

Tuesday 5[th] January, 2016

**Abstract**

In this report, we present the theory on aerodynamics of quadrotors using the well established momentum and blade element theories. From a robotics perspective, the theoretical development of the models for thrust and horizontal forces and torque (therefore power) are carried out in the body fixed frame of the quadrotor. Using momentum theory, we propose and model the existence of a horizontal force along with its associated power. Given the limitations associated with momentum theory and the inadequacy of the theory to account for the different powers represented in a proposed bond graph lead to the use of blade element theory. Using this theory, models are then developed for the different quadrotor rotor geometries and aerodynamic properties including the optimum hovering rotor used on the majority of quadrotors. Though this rotor is proven to be the most optimum rotor, we show that geometric variations are necessary for manufacturing of the blades. The geometric variations are also dictated by a desired thrust to horizontal force ratio which is based on the available motor torque (hence power) and desired flight envelope of the vehicle. The detailed aerodynamic models obtained using blade element theory for different geometric configurations and aerodynamic properties of the aerofoil sections are then converted to lumped parameter models that can be used for robotic applications. These applications include but not limited to body fixed frame velocity estimation and individual rotor thrust regulation [1, 2].

# 1  Introduction

In recent years, there has been an increased interest within the robotics community in understanding the aerodynamics of quadrotors to enhance modelling and control of such platforms. The only available references are those written for helicopters by aerodynamicists. Some of these references include [6, 9, 17, 21] and contain theories developed for aerodynamicists and helicopter performance analysts. From a robotic point of view, many of the parameters in the theories are immeasurable and therefore not available for control purposes. Hence the arguments in the report are driven by robotic and not aerodynamic arguments. In addition, the non-linear scaling of the forces and their effects and mechanisms (e.g. Bell-Hiller system), blade geometries and disc loading are different between quadrotors and full-sized helicopters. Quadrotors are designed based on simplicity, ease of maintenance and cost. For these reasons, the majority of quadrotors have fixed pitched blades contrary to the variable pitch and flapping hinges of the rotor blades typical of helicopters. An example of a helicopter rotor mechanism is shown in Figure 1. In this report, we present the relevant aerodynamic models for quadrotors based on the well established momentum and blade element theories developed primarily for helicopters. From these theories, we extend our analysis to produce simplified or lumped aerodynamic models that can be used in robotic applications. The quadrotor used in the analysis is that used in [5] which weighs $1.2kg$ and has $10in$ diameter propellers.

The report starts by describing the different frames of reference on quadrotors and their rotor blades. By choosing the body fixed frame, Section 2 uses momentum theory to model thrust and horizontal force and torque. By looking at a bond graph representation of power and axial flights, the limitations of momentum theory which include its inability to model some of the vortex states and a distribution of the forces across the rotor implies that another modelling technique has to be used. In Section 3 to 6, blade element theory is applied to individual elements to produce models for the forces and torque for different blade geometries and aerodynamic properties of the aerofoil sections used on quadrotors. In Section 3, the modelling framework for elemental forces and torque is developed. With the assumption of constant chord and pitch, infinite *aspect ratio* (AR) and zero-lift angle of attack, which represent the simplest rotor blades used predominantly in the helicopter literature, lumped parameter models for thrust, horizontal force and torque (hence power) are developed in Section 4. In Section 5, the assumptions of zero-lift angle of attack aerofoil and infinite *aspect ratio* are removed and a more complicated model is developed. In Section 6, ideal twist and chord are considered in developing models for the so-called ideal rotors used on the majority of quadrotors. In the development of the models for the ideal rotor, we show that the geometry is a design parameter that must be optimised with respect to the flight envelope of the quadrotor and the mechanical properties of the material used in manufacturing the blades.

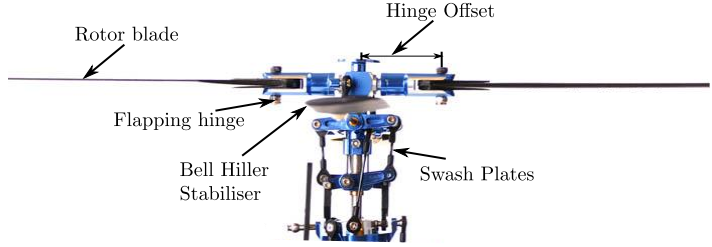

Figure 1: A helicopter rotor mechanism with Bell-Hiller flybar and swash plates mechanism for controlling cyclic and collective pitch.

## 2 Momentum Theory of Rotors

In this section, we present the aerodynamics of the rotor blades on quadrotors using momentum theory. The models for thrust, horizontal force and power are derived with the necessary assumptions contained in the theory. Starting with frames of reference, we carry out detailed momentum theory analysis of quadrotor rotor blades. To account for the fact that we are modelling in the body fixed frame of the quadrotor and not the rotor plane used in helicopter literature, we propose the existence of a horizontal force with an associated required power. The section concludes with a bond graph representation of the different powers and the limitations of momentum theory.

### 2.1 Frames of Reference

Consider Figure 2 which shows a quadrotor along with a rotor and the different planes and frames of reference. We consider that there exists a fixed frame on the Earth's surface termed inertial frame $\{A\}$. Attached to the quadrotor which is assumed rigid is the body fixed frame $\{B\}$. If we let the relative velocity of $\{B\}$ to $\{A\}$ be $V \in \{B\}$ where $V \in \mathbb{R}^3$. If there is wind blowing at a velocity of $W \in \mathbb{R}^3$ relative to $\{A\}$, then the total air relative velocity seen by the quadrotor is $-W + V$ expressed in $\{B\}$. Throughout the report $\vec{e}_1, \vec{e}_2, \vec{e}_3 \in \mathbb{R}^3$ are used to denote unit vectors in $x, y$ and $z$ directions respectively.

Taking a closer look at the rotor shaft and rotor plane, the rotor shaft is always aligned with the $\vec{e}_3$ axis of $\{B\}$. This implies that the hub reference frame and the body fixed frame $\{B\}$ are equivalent. We define the rotor reference frame $\{C\}$ which has its $\vec{e}_3$ aligned with that of $\{B\}$ and the $\vec{e}_1, \vec{e}_2$ directions in rotation at the speed of the rotor ($\varpi$) relative to $\{A\}$. In the sequel, the disposition of $\vec{e}_1$ of $\{C\}$ from that of $\{B\}$ is referred to as the azimuthal angle $\psi$. For the rotating rotor, a plane on its spinning tips is referred to as the tip path plane denoted $\{D\}$ and is otherwise known as the axis of zero flapping. In addition, it is the plane on which an observer does not see the conning or tilting of the rotor disc. It is a known phenomenon that as the spinning rotor translates, it tilts sideways and backwards and forms a cone around the rotor hub (shown later in Figure 8) thus moving this plane ($\{D\}$) above the rotor hub and putting it at an angle to the hub and therefore $\{B\}$. This phenomenon is referred to as blade flapping and in the sequel will be modelled by the angle $\beta$. If the quadrotor has variable pitch propellers such as helicopters, there is an additional plane known as the *no-feathering plane* $\{NFF\}$ and is normal to the plane of the swash plates. To minimise the effect of blade flapping, quadrotors are designed with most often fixed pitch rotor blades that are very stiff and rigid. For consistency with current quadrotor dynamic models, the analysis and theoretical developments will be carried out in the rotor hub frame which aligns with the quadrotor body fixed frame $\{B\}$. It should be noted that derotating $\vec{e}_1, \vec{e}_2$ of $\{C\}$ gives the $\vec{e}_1, \vec{e}_2$ of $\{B\}$ through the azimuthal angle ($\psi$) and the flapping angle ($\beta$) which depends on the rotor azimuthal position as such is denoted by $\beta(\psi)$. This is the approach taken in the blade element theory (Sections 3 to 6) in developing models for the different forces, torque and power for the different rotor geometric and aerodynamic configurations. The frames $\{C\}$ and $\{D\}$ along with $\beta(\psi)$ will also be explained and modelled in detail in the next sections.

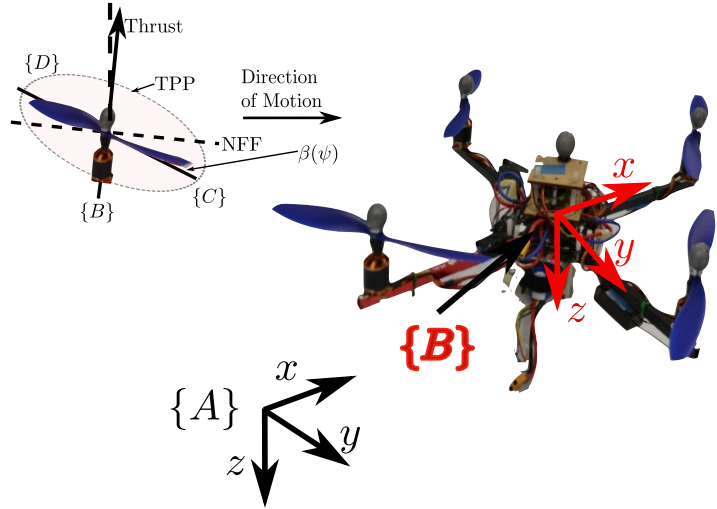

Figure 2: A quadrotor and the different frames of reference. In addition to $\{A\}$ and $\{B\}$ defined for quadrotors, the rotor and rotor hub also have additional frames of reference which include the rotor reference frame $\{C\}$, the tip path plane $\{D\}$ and the no-feathering plane $NFF$. These are also shown in the figure.

**Remark 1** *We choose to model the thrust force $T \in \mathbb{R}$ in the $\vec{e}_3$ direction of $\{B\}$. It is the component of force along the rotor shaft/hub used in the quadrotor dynamic modelling literature [4, 5]. As the rotor translates, a force perpendicular to $T$ lying on the plane containing $\vec{e}_1, \vec{e}_2$ in $\{B\}$ is generated to oppose the motion. This force we model by the so-called in-plane horizontal force ($H \in \mathbb{R}^3$). In common practice for slow moving quadrotors, $H^\top \vec{e}_3 = 0$. As the rotor translates, there is tilting of the* tip path plane *($\{D\}$) from the rotor hub/shaft as a result of blade flapping and this causes the misalignment of the* tip path plane *and the rotor shaft ($\{B\}$). This creates one of the components of the H-force. In addition, there is an induced drag term associated with forward motion. Blade flapping and induced drag have been lumped in the quadrotor literature and used to improve the control performance of multirotor vehicles [13].*

**Remark 2** *For helicopters, thrust is changed by using the feathering mechanism to change the blade pitch either collectively or cyclically using a swash plate and the Bell-Hiller mechanism shown in Figure 1. For the majority of small-scale electrically powered fixed pitch quadrotors, thrust changes are achieved by changing the speed of the rotor using an* electronic speed controller *(ESC). This is because quadrotors have a low moment of inertia blades relative to helicopters that make rotor speed changes easily achievable.*

## 2.2  Introduction to Momentum Theory

Momentum theory for rotary wing vehicles was developed by Glauert based on earlier work by Froude for aircraft propellers. It is one of the two most popular theories for propeller analysis. Its simplistic approach has made it the starting point for modelling aerodynamic forces on rotors. The theory considers a rotor as an actuator disc with the accelerating air forming a streamtube. The control volume shown in Figure 3 shows this streamtube. As the air is sucked and accelerated by the rotor as it goes through it, it generates a virtual induced airflow with velocity $v^i \in \mathbb{R}^3$ or $v^i = (v_x^i, v_y^i, v_z^i)^\top$. Glauert made several assumptions which are stated in Assumption 2.1.

**Assumption 2.1** *These assumptions include*

- *The rotor disc has an infinite number of rotor blades such that there is a uniform constant distribution of aerodynamic forces over the rotor disc.*

- *The rotor disc is an infinitely thin disc of area A which offers no resistance to air passing through it.*
- *The flow is irrotational and therefore no swirl is imparted to it.*
- *The air outside the streamtube remains undisturbed by the actuator disc.*

It has been observed that the higher the disc loading, the more these assumptions hold. The disc loading (DL) is defined as

$$DL = \frac{T}{A}.$$

Noting the quadrotor considered in this report (and quadrotors in general) has a higher disc loading than the majority of helicopters, indicates that momentum theory holds better for quadrotors than for helicopters.

**Remark 3** *As we are considering the entire rotor disc, the forces generated are in the rotor hub plane, i.e. $\{B\}$, and any effects resulting from blade flapping are not considered in the control volume but should be seen as a validation of the existence of the proposed H-force. Details on blade flapping and its effects on the generated forces and torque will be covered in detail in Section 3. It should be noted that blade flapping is responsible for the misalignment of the rotor plane and the rotor hub shown in Figure 4.*

## 2.3 Momentum Theory for z-Direction or Axial Motion

To develop the thrust and power models using momentum theory, consider the simplest of control volumes shown in Figure 3 where the velocity of the vehicle is $V = (0, 0, V_z)^\top$. It should be noted that in $\{B\}$, $V_z > 0$ indicates that the vehicle is moving downwards based on our right hand coordinate system shown in Figure 2. For the rotor in hover or undergoing purely axial motion, we make the following additional assumptions

**Assumption 2.2** *The flow through the rotor is one-dimensional, quasi-steady, incompressible, inviscid, and behaves as an ideal fluid and the radius of a plane perpendicular to the control volume at the rotor disc equals the rotor radius.*

From Froude's theory, the airflow through the propeller disc is continuous and characterised by a constant speed which at hover is the induced velocity denoted by $v^i$. The propeller disc introduces a discontinuity in the pressure. This discontinuity is denoted by $\Delta P$ and can also be thought of as the increment in the static pressure across the disc. As shown in Figure 3, we consider also a virtual cylindrical surface of radius $R_0 > R$ containing the propeller disc of radius $R$ and displaced along the propeller spin axis. This surface will be employed only to compute the amount of air flowing inside and outside the propeller disc. More specifically, let us denote as *upstream plane* the disc of radius $R_0$ and as *downstream plane* the disc of radius $r_2 < R$ which are located at the beginning and at the end of the cylinder respectively. By assuming that they are infinitely far from the propeller disc implies that the streamlines are parallel to the propeller spin axis. By applying momentum theory, the thrust force $T$ (in the axial direction i.e. $\vec{e}_3$ of $\{B\}$) can be computed as the difference between the momentum of the flux going out and the momentum of the flux coming into the streamtube i.e.

$$\begin{aligned} T &= \dot{m}(V_z^\infty - V_z) - \dot{m}(-V_z), \\ &= \dot{m} V_z^\infty, \end{aligned} \qquad (1)$$

where $\dot{m}$ is the mass flow rate which is defined as

$$\dot{m} = \rho A |V^a|,$$

and

$$V^a = \begin{pmatrix} 0 \\ 0 \\ v_z^i - V_z \end{pmatrix},$$

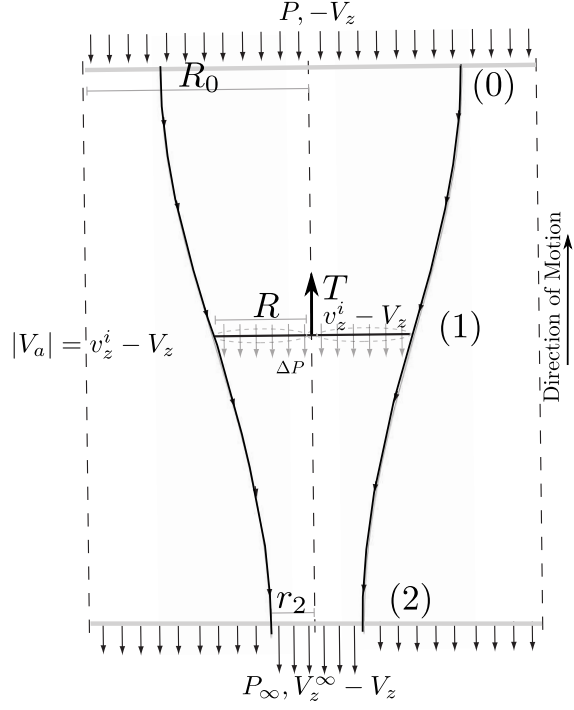

Figure 3: Vertical streamtube for hover and axial analysis. In the figure, there are three sections: *upstream* (0), rotor disc (1) and *downstream* (2) to help with the analysis. The figure also shows the velocities at each of these sections. Please note for ease of analysis, $V^\infty$ is pointing downwards and therefore positive in $\{B\}$.

is the total air velocity at the rotor disc within the streamtube. The area of the streamtube is that of the rotor disc given by $A = \pi R^2$. The power supplied can then be computed as the product of the thrust and the local velocity across the disc (i.e., $v_z^i - V_z$ in the direction of $T$ along $\vec{e}_3$) and is given by

$$P_T = T(v_z^i - V_z). \tag{2}$$

This power is also the rate of kinetic power imparted into the air across the streamtube and is given by

$$P_T = \frac{1}{2}\dot{m}(-V_z + V_z^\infty)^2 - \frac{1}{2}\dot{m}(V_z)^2,$$

$$P_T = \frac{1}{2}\dot{m}(V_z^2 - 2V_z V_z^\infty + (V_z^\infty)^2) - \frac{1}{2}\dot{m}(V_z)^2.$$

Hence,

$$P_T = \frac{1}{2}\dot{m}\left((V_z^\infty)^2 - 2V_z V_z^\infty\right). \tag{3}$$

Substituting for $T$ using (1) in the power equation (2), and comparing it to (3), we get

$$\dot{m} V_z^\infty V^a = \frac{1}{2}\dot{m}\left((V_z^\infty)^2 - 2V_z V_z^\infty\right).$$

This implies that

$$v_z^i = \frac{1}{2}V_z^\infty.$$

Thus substituting for $\dot{m}$ and $v_z^i$ in (1), the aerodynamic thrust and power in the air for a given rotor in axial or $z$-direction flight are given by

$$T = 2\rho A v_z^i (v_z^i - V_z), \tag{4}$$

$$P_T = T(v_z^i - V_z). \tag{5}$$

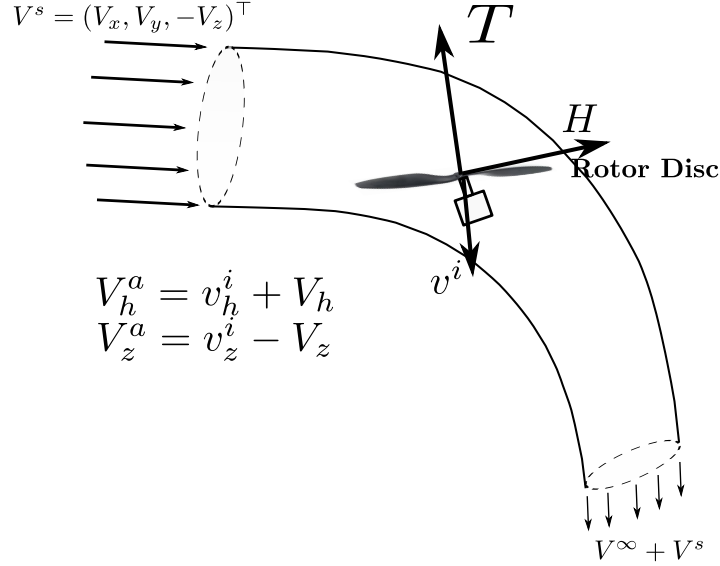

Figure 4: Rotor control volume for generalised motion. The figure shows the streamtube and generated forces, the velocity of the rotor $V$ and the resultant air velocity $V^a$ seen by the rotor along with the induced velocity at the disc. Please note for ease of analysis, $V^\infty$ is pointing downwards and therefore positive in $\{B\}$.

**Remark 4** *It is worth noting that at hover ($V_z = 0$), using (4) and (5), the known static relationship between thrust $T$ and power $P$ holds and is given by*

$$P = \frac{T^{\frac{3}{2}}}{\sqrt{2\rho A}}.$$

*In designing helicopters, large rotor blades are used since for the generation of the same thrust, there is less power required thus implying that helicopters have low disc loading. This equation can also be applied to the design of heavy lifting quadrotors. However as will be shown later, larger and heavier blades are disadvantageous for fast and agile quadrotor vehicles due to reduced transient performance of the rotor speed as a result of rotor mass moment of inertia. Hence high performance quadrotors have rotor blades with higher disc loading than helicopters.*

## 2.4 Momentum Theory for Generalised Motion

Consider now Figure 4 which shows a slightly tilted actuator disc to that of Figure 3. In this case, the rotor experiences both translational and vertical airflow. As such the induced airflow now has all the components in $x, y, z$. Figure 4 also shows a well known phenomenon of rotor blades, blade flapping which is responsible for the coning and backwards tilting of rotor blades. Its net effect is to create/increase any force in the plane of $\vec{e}_1, \vec{e}_2$ opposing the motion of the rotor. As a result of the coning and tilting, the $\vec{e}_3$ direction of $\{B\}$ and $\{D\}$ are misaligned. Hence the induced airflow $(v^i)^\top \vec{e}_3 \in \{D\}$ now has components in the horizontal plane and vertical direction of $\{B\}$. In the helicopter literature $v^i_h = (v^i_x, v^i_y, 0)^\top$ is ignored as it is relatively small compared to $v^i_z$ and its existence is as a result of $V_h$ and exists only in $\{B\}$ and not in $\{D\}$. This is because they assume a completely vertical flow and carry out their analysis in the tip path plane $\{D\}$. As shown previously in Section 2.3, for a purely vertical flight, there is only $v^i_z$. However, for a purely horizontal flight, there are $v^i_z$ and $v^i_x, v^i_y$.

**Remark 5** *In the hub frame ($\{B\}$), the rotor experiences the resultant horizontal/planar velocity. This velocity is represented by the subscript h. For example $V_h = (V_x, V_y, 0)^\top \in \{B\}$ is seen as the resultant $|V_h|$. Hence in the sequel, derivations are for a 2-D flow. To minimise notational confusion, we use $V_h, v^i_h, H$ to represent $|V_h|, |v^i_h|, |H|$ respectively.*

From Figure 4 and more obvious in Figure 5, it is quite clear that there is an additional force to $T$. This is the force ($H \in \mathbb{R}^3$) in the plane of the rotor acting against the motion of the plane that is perpendicular to the rotor hub. This force exists as we are modelling in the rotor hub or $\{B\}$ of the vehicle and not in $\{D\}$ as in traditional helicopter literature [6, 9, 21]. Before continuing with the analysis, we make the following assumption on the wind velocity $W$.

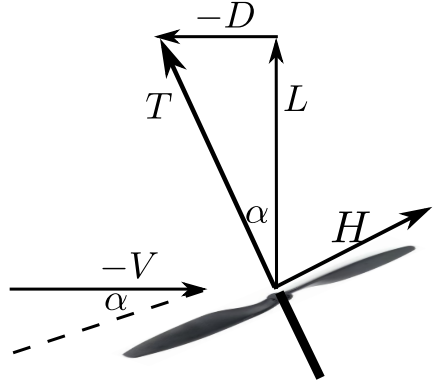

Figure 5: Rotor reference frame {C}. The figure shows the resolution of $T$ into components parallel and perpendicular to the airflow thus clearly showing the existence of $H$ as it is required for steady state flight.

**Assumption 2.3** *The wind velocity* $|W| = 0$.

It is however not tedious to incorporate the wind velocity $W$ into the computations by setting the air velocity to $(-W_x + V_x, -W_y + V_y, W_z - V_z)^\top$.

We reapply the same analysis as the axial flight in Section 2.3 and carry out the analysis in the rotor hub or $\{B\}$. It should be noted that because our $\vec{e}_3$ is pointing downwards, we introduce a new variable $V^s = (V_x, V_y, -V_z)^\top$ to represent the velocity at the farstream seen by the rotor. Starting with the application of momentum theory in the direction of motion of air around the rotor hub or $\{B\}$ by first considering the momentum of the fluid entering and leaving the control volume and letting the forces be $F = (H_x, H_y, T)^\top$ with $H^\top \vec{e}_3 = 0$ for slow moving vehicles, then

$$F = \dot{m}(V^\infty + V^s) - \dot{m}(V^s).$$

Hence,

$$F = \dot{m}V^\infty.$$

Taking note of the $\vec{e}_3$ of $V \in \{B\}$ and by letting

$$V^a = \begin{pmatrix} v_h^i + V_h \\ v_z^i - V_z \end{pmatrix},$$

be the total velocity of the air at the rotor hub, the power to generate this force is

$$P = F^\top V^a. \tag{6}$$

To determine the relationship between $V^\infty$ and $v^i$, it is worth noting that power is a scalar as such we can consider it as a result of horizontal and vertical motion separately and sum them to get the total power. Consider first the axial direction ($\vec{e}_3$) which contains the thrust force $T$. Applying Newton's second law or the conservation of momentum at the farstream and downstream and along the rotor hub ($\{B\}$),

$$F\vec{e}_3 = T = \dot{m}(V_z^\infty + (-V_z)) - \dot{m}(-V_z),$$
$$= \dot{m}V_z^\infty, \tag{7}$$

and the power ($P_T$) for generating this thrust is

$$P_T = T(v_z^i - V_z). \tag{8}$$

Power is also the rate of kinetic power imparted into the air across the streamtube. This is given by

$$P_T = \frac{1}{2}\dot{m}(V_z^\infty - V_z)^2 - \frac{1}{2}\dot{m}(-V_z)^2,$$

$$P_T = \frac{1}{2}\dot{m}(-2V_z V_z^\infty + (V_z^\infty)^2). \tag{9}$$

Comparing (8) and (9) and substituting for $T$ using (7), we get

$$\dot{m}V_z^\infty(v_z^i - V_z) = \frac{1}{2}\dot{m}(-2V_z V_z^\infty + (V_z^\infty)^2).$$

Hence after expansion and cancellation of the $V_z^\infty$ terms as in Section 2.3, we get

$$v_z^i = \frac{1}{2}V_z^\infty.$$

Substituting for $V_z^\infty$ and $\dot{m} = \rho A V^a$ in (7), we obtain

$$T = 2\rho A v_z^i |V^a|. \tag{10}$$

The same relationship for the induced and downstream velocity components for the horizontal motion can be obtained

$$v_h^i = \frac{1}{2}V_h^\infty.$$

Thus in a similar manner to $T$, the horizontal force $H$ generated that is acting against the direction of motion of the plane at the rotor hub and power in generating this force are given by

$$H = 2\rho A v_h^i |V^a|,$$

and

$$P_h = H(v_h^i + V_h),$$

respectively. The original generalised momentum theory equations which do not include the $H$ force and $v_h^i$ though were originally developed for vertical or axial flights as shown in Section 2.3 have not been theoretically verified. There are however experimental evidence supporting the theory [21, pg. 51-52]. In the literature, the $H$ force is the drag force and it has been used in the estimation of the body fixed frame velocity of quadrotors [1, 2]. The modified generalised momentum theory equations are presented in (11) to (14).

$$T = 2\rho A v_z^i |V^a|, \tag{11}$$

$$H = 2\rho A v_h^i |V^a|, \tag{12}$$

$$P = T(v_z^i - V_z) + H(v_h^i + V_h), \tag{13}$$

$$V^a = \begin{pmatrix} v_h^i + V_h \\ v_z^i - V_z \end{pmatrix}. \tag{14}$$

In Figure 6, we show the different powers generated and consumed at the rotor hub. In the diagram $P_p$ represents the profile power dissipated as a result of the spinning of the rotor blades. It should be noted that momentum theory does not account for this power and will be later modelled in Section 3. Intuitively, profile power can be seen as $P_p = \tau_a \varpi$, where $\tau_a$ is the air resistance torque and $\varpi$ the speed of the rotor. From experiments performed in [3, 4], the aerodynamic power (13) can be rewritten as

$$P_a = \frac{1}{\text{FoM}}(P_T + P_h),$$

$$= \frac{1}{\text{FoM}}\left(T(v_z^i - V_z) + H(v_h^i + V_h)\right),$$

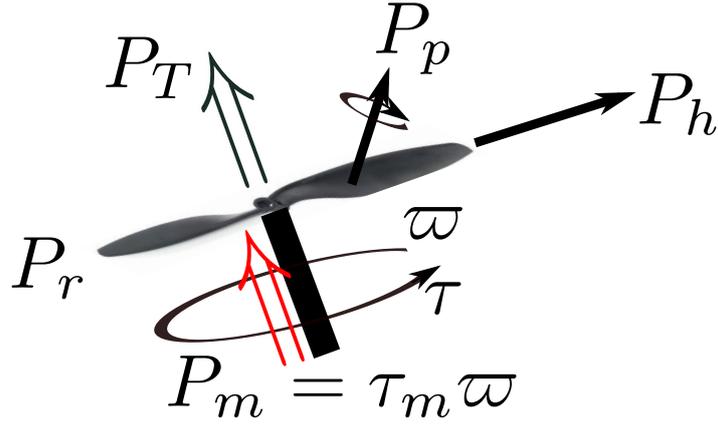

Figure 6: The different powers on a spinning rotor.

where $FoM$ is the *figure of merit* and is defined as an efficiency factor to account for non-ideal losses and the region not modelled by momentum theory as shown in the bond diagram in Figure 7. If $P_m$ is the mechanical power supplied to the rotor shaft, then the $FoM$ is defined as

$$FoM = \frac{TV^a}{P_m}. \tag{15}$$

In helicopter literature, losses occur in two parts: rotor wake and tip vortices. Typical values are 10% for wake flow and 15% for tip vortices and other losses [9, pg. 45-46]. From experiments performed in [3, 4], the $FoM$ was found to be between 60 and 70% for most rotors. The theoretical maximum of $FoM$ at hover is 81% [9, pg. 47] though values as high as 77% have been recorded [15, 16]. It should be noted that the $FoM$ is the same for both axial and planar axis because the properties of the fluid are uniform and so the non-ideal loses are the same in every direction.

Carrying out a power balance at the rotor hub and ignoring the power due to profile blade losses and if $P_r$ is the power dissipated in accelerating the rotor to a constant rotor speed ($\varpi$), the mechanical power $P_m$ defined as the power supplied to the rotor shaft is given by

$$P_m = P_r + P_a.$$

In reality, this is the power that is controlled and can be set to a desired value. The $P_r$ is estimated using methods described in [3]. A bond diagram showing the power flow from the rotor shaft to the air and the generated forces on the vehicle are shown in Figure 7. Starting from the rotor shaft which acts as a source of effort producing for the entire vehicle the torque $\tau$ and rotor speed $\varpi$ with the power lost through the resistance $R_l = P_r$ modelled by a 1-junction. Through algebraic equations i.e. a modulated transformer MTF, the distribution of forces $f(r,\psi)$ and velocity $U(r,\psi)$ are obtained. These then go through a 0-junction and an MTF to produce the force $F$ and velocity $V^a$ with power losses in the wake represented by the conductance $\frac{1}{R_{l1}}$. Applying momentum theory to this region through the 0-junction, the power used in moving the vehicle ($FV$) and power lost in the induced flow ($Fv^i$) are seen through the conductance $\frac{1}{R_{l2}}$. The distribution of forces $f(r,\psi)$ and velocities $U(r,\psi)$ will be dealt with using blade element theory which is covered in the remainder of the report.

## 2.5 Rotor Wake and Vorticity

The induced velocity distribution is not constant across the rotor as would be expected especially for forward flights where its value at the leading edge (face directed into wind) is greater than at the trailing edge (face away from wind). To account for such a non-constant induced velocity distribution, let $v^{i0} \in \mathbb{R}^3$ be the mean induced

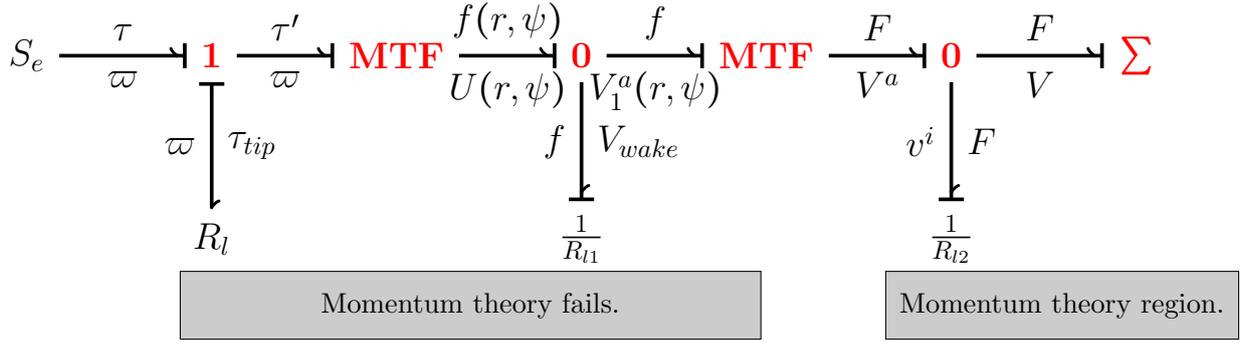

Figure 7: Bond diagram for a motor/rotor system. The source of effort ($S_e$) is the motor shaft which makes the vehicle to fly and overcome vortices and wake. The figure also shows that the forces on a rotor should be modelled by a distribution $f(r,\psi)$ and velocity $U(r,\psi)$ at every radial position and azimuthal angle. In addition the figure also shows losses represented by conductance and resistances ($R_l$) which are embedded in the $FoM$.

velocity at the disc centre and its components are given by (based on (11) and (12) respectively)

$$v_z^{i0} = \frac{T}{2\rho A |V^a|},$$
$$v_h^{i0} = \frac{H}{2\rho A |V^a|}.$$

Glauert proposed the following induced velocity model

$$v_z^i = v_z^{i0}\left(1 + \kappa_e \frac{r}{R}\cos\psi\right),$$

where $r \in [0,R]$ is the radial position, $\psi$ is the azimuthal angular position and $\kappa_e$ is a number greater than unity (usually 1.2 for helicopters) [21, pg. 54]. For our modelling application, we believe that this model also holds for the proposed $v_h^i$ with the same $\kappa_e$ value. Details on a better induced velocity model are presented in Section 3.4 using the Mangler and Squire's method. A drawback of these models is the $v^{io}$ dependence on itself through $V^a$ and on thrust both of which are unknown for small-scale open-source quadrotors. From a robotics perspective, the induced power (power associated with induced velocities) in the power model (13) can be modified to account for the non-uniform and non-constant induced velocities. Thus the modified power equation is

$$P = T(\kappa_e v_z^i - V_z) + H(\kappa_e v_h^i + V_h).$$

From static tests performed in [3, 4], without any exact measurements of $V^a$, it can be said that $\kappa_e$ is embedded in the $FoM$ value obtained. Hence proper wind tunnel tests where accurate measurements of $V, V^a$ are necessary to experimentally determine $\kappa_e$ and $FoM$. In the sequel, it will be shown that unlike helicopters, quadrotor blades are designed to produce substantial amount of lift across the different sections of the blades and not only at the outer blade sections as a result of blade geometry which suffer from significant tip vortices in the case of helicopters.

### 2.6 Limitations of Momentum Theory

There are flight regimes in which momentum theory fail. These regimes are associated with axial motion corresponding to cases where the streamtube model and Assumptions 2.1 and 2.2 are no longer valid. These axial flight regimes which rotors of quadrotors and helicopters experience in addition to the *normal* state are outlined below [4].

**Vortex Ring State (*VRS*):** The rotor is said to be in this state when the rate of descent is half the vertical induced velocity. In this state, vortex ring encircles the disc causing the flow to become unsteady resulting in high levels of vibration. It should be noted that momentum theory cannot be used to model in this state.

**Turbulent Wake State (*TWS*):** This state occurs when the rate of descent of the rotor equals that of the vertical induced velocity as such there is no net flow of air through the rotor. From (11) and (13), it is easily seen that there is no thrust generated and the power required is zero. This contradicts the fact that power has to be supplied to maintain the spinning of the rotors. Thus indicating that momentum theory cannot be applied. However, the vibrations are less in this state than those of the *VRS*.

**Windmill Brake State (*WBS*):** This state occurs when the rate of descent of the rotor is more than twice the induced velocity. At this rate, the blade sections are likely to stall. In this state, the net flow of air is entirely upwards thereby creating negative thrust which causes a consumption of power from the air. Of the three axial descent states mentioned, momentum theory models for $T$ and $P$ are only valid in this state [4],[9, pg. 60], [21, pg. 10-13].

Unlike axial flights, there are no limitations caused by the horizontal/planar motion as both $|v_h^i|$ and $|V_h|$ are always positive in the direction of motion. Contrary to helicopters, to avoid such axial states for quadrotors, more power is supplied to the motors. By doing so, the vortices on the tip of the rotor blades are blown away. In the subsequent sections, it will be shown that most quadrotor blades are designed to have optimal chord with the entire blade span generating lift and drag forces required to generate $T, H$ and $P$. Hence the tip vortices have minimal effects on the aerodynamic forces generated compared to helicopters which have the majority of the aerodynamic forces generated in the outer portion of the blade.

In the next sections, blade element theory is presented. It uses an element of a blade to model forces and torque (hence power) irrespective of the axial flight condition. In addition, it makes use of the fact that the elemental forces and velocities along the elements of a rotor are functions of the elemental radial and azimuthal position.

## 3  Introduction to Blade Element Momentum Theory

Blade element theory considers the individual elements of a rotor blade, the aerodynamic properties (lift and drag coefficients) of the aerofoil, blade geometry and uses elemental forces and torque. The overall model for thrust ($T$), in-plane horizontal force ($H$) and torque ($\tau$) and power ($P$) are obtained by integrating along the entire blade and over a rotor revolution of these elemental forces and torques. In this section, we examine the necessary assumptions and models for induced velocity and blade flapping that are used in the development of the elemental forces and torque. These models are then incorporated into the model for the total velocity along with its horizontal and vertical components. This then led to the examination of the different elemental aerodynamic forces in the tip path plane $\{D\}$ and $\{C\}$ and their resultants which is the thrust and horizontal force in the body fixed frame $\{B\}$. The final elemental results will be used in the subsequent sections to model $T, H$ and $P$ for a variety of rotor geometries and aerodynamic properties.

The following assumptions form the basis of blade element theory.

**Assumption 3.1** *The outward centripetal force acting on the blades ensure that they can be assumed rigid and do not stall.*

It should be noted that similar to Section 2, the development of the models for $T, H$ and $P$ are carried out in the rotor hub frame or body fixed frame $\{B\}$. The airflow consists of the vertical component and a horizontal component which is the resultant of the airflow in the $\vec{e}_1$ and $\vec{e}_2$ directions. For this reason, we only model the thrust $T$ acting in $\vec{e}_3$ and the magnitude of the horizontal force $H = |H|$ in the plane containing $\vec{e}_1, \vec{e}_2$.

### 3.1  Theoretical Framework

Since blade element theory considers individual elements of the rotor, it is necessary to obtain the total of these elements along the span of the rotor which defines the physical quantity of interest. In addition, the rotor spins resulting in azimuthal changes from 0 to $2\pi$. Hence we can model the elemental contribution of a quantity say force $F$ as $\mathrm{d}F(r, \psi)$ for an element located at a distance $r$ from the rotor hub ($r \in [0, R]$) and $\psi$ is its azimuthal angular displacement. The total force is then given by the sum of the individual elements across all the azimuthal positions for the blade. For $N_b$ number of blades, the total force is given by

$$F = \frac{N_b}{2\pi} \int_0^{2\pi} \int_0^R dF(r,\psi) \mathrm{d}r \mathrm{d}\psi. \tag{16}$$

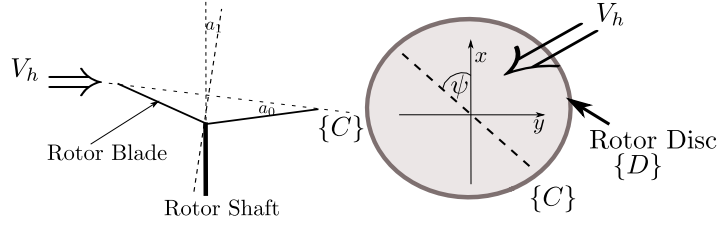

Figure 8: A view of the rotor-shaft system of quadrotors. The figure shows a side view with the coning angle $a_0$ and sideways tilt $a_1$ relative to $\{B\}$ which are part of the flapping angle $\beta$. In addition, there is the top view which shows the TPP relative to the $\vec{e}_1$ of $\{B\}$ and its disposition, the azimuthal angle $\psi$.

**Remark 6** *From Section 2.5, it was pointed out that rotors shed tip vortices which are accounted for by using a tip loss model. If $c_t$ is the chord of the rotor at the tip, if $B \in [0,1]$ where $BR$ is the effective radius of the rotor, then $B$ is defined by the Prandtl tip loss function*

$$B = 1 - \frac{c_t}{2R}.$$

*Glauert used this to model tip loss such that the radial limits of the integrand (16) are $[0, BR]$. In the sequel $B = 1$ is used to obtain a simplified model. $BR$ will be a coefficient to be determined experimentally as will be shown in the lumped parameter model presented in Section 4.*

## 3.2 Relevant Definitions

Before presenting blade element theory, it is worth defining some non-dimensional variables in line with Remark 1 which states that the forces and torque are modelled in $\{B\}$ i.e. along and perpendicular to the rotor shaft/hub. If $I_b$ is the mass moment of inertia of the blades, $N_b$ is the number of blades, $c$ is the chord length of the blade, $R$ is the blade radius and $\varpi$ is the speed of the rotor, then

**Definition 1** *The rotor hub/shaft advance ratio $\mu$ in the direction of $V_h$*

$$\mu = \frac{|v_h^i + V_h|}{\varpi R} = |\mu^i + \mu_h|.$$

**Definition 2** *The rotor hub/shaft vertical inflow ratio $\lambda$ in the $\vec{e}_3$ of $\{B\}$*

$$\lambda = \frac{v_z^i - V_z}{\varpi R} = \lambda^i - \lambda_z.$$

**Definition 3** *The Solidity ratio*

$$\sigma = \frac{N_b c}{\pi R}.$$

**Definition 4** *The Lock number for a constant chord blade*

$$\gamma = \frac{\rho A c R^4}{I_b}.$$

We define two further coefficient variables which will be shown in the sequel to be dependent on the aerodynamic state of the rotor.

**Definition 5** *The thrust coefficient*
$$C_T = \frac{T}{\varpi^2}.$$

**Definition 6** *The power coefficient*
$$C_P = \frac{P}{\varpi^3}.$$

## 3.3 Model for Blade Flapping Angle ($\beta(\psi)$)

With the assumption of steady state flight, we model the blade flapping angle $\beta$ at an azimuth $\psi$ using the following Fourier series expression with harmonic terms $(a_0, a_1, \ldots, a_n, b_1, \ldots, b_n)$ by

$$\beta(\psi) = a_0 - \sum_{n=1}^{\infty} a_n \cos n\psi - \sum_{n=1}^{\infty} b_n \sin n\psi. \tag{17}$$

Further details on the treatment of blade flapping as a drag term is presented in [4].

In this work, we consider only the first harmonics, i.e. $n = 1$, since the higher harmonic terms $a_2, b_2, a_3 \ldots$, can be ignored as they are very small or negligible for short rigid rotors. The blade flapping model is thus given by

$$\beta(\psi) = a_0 - a_1 \cos \psi - b_1 \sin \psi. \tag{18}$$

The derivative with respect to $\psi$ is given by

$$\frac{\mathrm{d}\beta(\psi)}{\mathrm{d}\psi} = a_1 \sin \psi - b_1 \cos \psi. \tag{19}$$

And the time derivative of $\beta(\psi)$ is

$$\begin{aligned}\dot{\beta}(\psi) &= \frac{\mathrm{d}\beta(\psi)}{\mathrm{d}\psi} \frac{\mathrm{d}\psi}{\mathrm{d}t}, \\ &= \varpi \frac{\mathrm{d}\beta(\psi)}{\mathrm{d}\psi}.\end{aligned} \tag{20}$$

From Figure 8, $a_0$ can be seen as the coning angle of the blade and $a_1$ and $b_1$ as the $-\vec{e}_1 \in \{B\}$ and $-\vec{e}_2 \in \{B\}$ tilt of the rotor disc or the tip path plane. The coefficient $a_0$ is strongly linked to blade rigidity and is very small for quadrotors since they have high rigidity and stiffness rotor blades. It should be noted that even at hover, there is an $a_0$ for long, slender and fully flexible rotor blades especially those used on helicopters. The presence of $a_0$ does not cause any misalignment of the *TPP* or $\{D\}$ from $\{B\}$. The other coefficients $a_1$ and $b_1$ are the backward and sideways tilting of the *TPP* from $\{B\}$ and therefore results in the misalignment of $\{D\}$ from $\{B\}$.

If $\theta_0$ is the blade pitch/twist, then the flapping coefficients for a no flapping hinge offset are defined by [6, pg. 107]

$$a_0 = \frac{\gamma}{8}\left[\theta_0 \left(1 + \mu^2\right) - \frac{4}{3}\lambda\right], \tag{21}$$

$$a_1 = \frac{2\mu \left(4\theta_0/3 - \lambda\right)}{1 - \mu^2/2}, \tag{22}$$

$$b_1 = \frac{\frac{4}{3}\mu a_0}{1 - \mu^2/2}. \tag{23}$$

Given that $\mu$ is small such that $\frac{1}{1\pm\mu^2/2}$ can be approximated to 1, the flapping angle coefficients are

$$a_0 = \frac{\gamma}{8}\left[\theta_0 \left(1 + \mu^2\right) - \frac{4}{3}\lambda\right], \tag{24}$$

$$a_1 = 2\mu \left(4\theta_0/3 - \lambda\right), \tag{25}$$

$$b_1 = \frac{4}{3}\mu a_0. \tag{26}$$

Multiples and higher powers of these coefficients are such that they can be neglected. This is used in the computations of the $H$-force and torque.

One of the root causes of blade flapping is the dissymetry in the lift during forward flights. This dissymetry creates a moment at the rotor hub. To minimise this moment, rotors are either connected to a hinge at the hub or rigidly attached and cyclically feathered by decreasing the pitch of the advancing and increasing the pitch of the retreating blade thus removing the lift imbalance (see Figure 1). Furthermore as was pointed out, quadrotors neither have flapping hinges nor swash plates but have short rigid and stiff blades to minimise the flapping effect.

## 3.4 Induced Velocity ($v^i(r,\psi)$) Distribution

In order to apply blade element theory, the induced velocity distribution along the span and different azimuth angles must be known. The estimation of the distribution of the vertical component of the induced velocity (and therefore proposed horizontal induced velocity) distribution on the rotor is a complex problem. In the theoretical development of the forces and torque, we will assume a constant $v^i$ distribution and therefore have the induced and translational velocities modelled through $\lambda$ and $\mu$. Furthermore, a good approximation for $v^i$ was proposed by Mangler and Squire [11]. The model treats the rotor as a lifting surface with a pressure jump. The induced velocity field is modelled as a small perturbation superimposed upon an otherwise uniform velocity field. The velocity distribution is expressed by a Fourier series of harmonic terms given by (27)

$$v^i(r,\psi) = 4v^{i0}\left[\frac{1}{2}d_0 + \sum_{n=1}^{\infty} d_n\left(r,\alpha_\mathrm{D}\right)\cos n\psi\right], \tag{27}$$

where $\alpha_\mathrm{D}$ is the rotor disc incidence and $v^{i0} \in \mathbb{R}^3$ is the mean induced velocity. To obtain $v^{i0}$, recall from Section 2 the total velocity of the air through the rotor

$$V^a = \begin{pmatrix} v_h^{i0} + V_h \\ v_z^{i0} - V_z \end{pmatrix},$$

where $V$ is the velocity of the vehicle. Using momentum theory, we restate the mean induced velocities are given by

$$v_z^{i0} = \frac{T}{2\rho A |V^a|}, \tag{28}$$

$$v_h^{i0} = \frac{H}{2\rho A |V^a|}. \tag{29}$$

This use of momentum theory in determining the mean induced velocity $v^{i0}$ is the reason the theoretical development is also referred to as blade element momentum theory ($BEMT$). The harmonic terms or coefficients $d_i$ are [6, pg. 82-83]

$$d_0 = \frac{15}{8}\eta\left(\frac{r}{R}\right)^2,$$

$$d_1 = -\frac{15\pi}{256}\left(5 - 9\eta^2\right)\left(\frac{r}{R}\right)\left(\frac{1 - \sin\alpha_\mathrm{D}}{1 + \sin\alpha_\mathrm{D}}\right)^{1/2},$$

$$d_3 = \frac{45\pi}{256}\left(\frac{r}{R}\right)^3\left(\frac{1 - \sin\alpha_\mathrm{D}}{1 + \sin\alpha_\mathrm{D}}\right)^{3/2},$$

where $\eta^2 = 1 - (r/R)^2$. For $n = 2k$, $k = 1, 2, \ldots, \infty$,

$$d_n = (-1)^{(n-2)/2}\frac{15}{8}\left[\frac{\eta + n}{n^2 - 1} \cdot \frac{9\eta^2 + n^2 - 6}{n^2 - 9} + \frac{3\eta}{n^2 - 9}\right]\left(\frac{1 - \eta}{1 + \eta}\right)^{n/2}\left(\frac{1 - \sin\alpha_\mathrm{D}}{1 + \sin\alpha_\mathrm{D}}\right)^{n/2}.$$

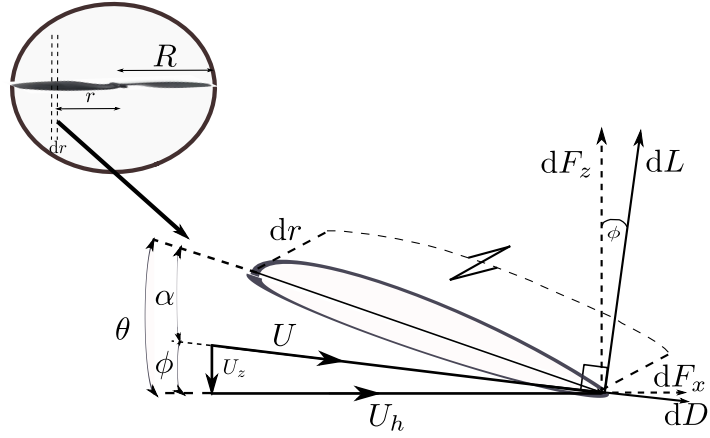

Figure 9: An aerofoil section of a blade at a location $r$ from the rotor hub. The figure also shows the different elemental forces which include lift and drag and the horizontal and vertical forces on the aerofoil in $\{D\}$. The different angles and velocity components are also shown.

For $n = 2k + 1, k \geq 2$,
$$d_n = 0.$$

As has been stated in Section 3.3, we are only concerned with the primary modes and therefore first harmonic flapping motion. However, $d_1 \ll d_0$ and $\alpha_D$ introduces another unknown which is not measured as it is a function of $\beta(\psi)$. The final simplified model for the induced velocity is given by

$$v^{\mathrm{i}}(r, \psi) = 4v^{\mathrm{i}0}\left(\frac{1}{2}d_0\right). \tag{30}$$

It will shown later in the sequel that (30) can be represented by a constant i.e. $v^i(r, \psi) = v^i$ and later on changed as it does not change the computations. Hence, the modelling process does not make use of the Mangler and Squire model for $v^i(r, \psi)$. By choosing a constant $v^i$ implies that we are not required to know the thrust before hand (through (28) and (29)) as for present day quadrotor technology, $T$ as well as the induced velocity $v^i$ are unknown. Furthermore, given the near hovering flight envelope for which quadrotors are designed and their use of ideal rotors (see Section 6) implies that this assumption is valid.

Similarly, the current helicopter literature, uses these assumptions in the development of the models for $T, H$ and $P$. A consequence of this is a model for a lower power as the induced power is underestimated but can however be compensated for during the calibration process and the introduction of a scaling factor in the final induced power model.

### 3.5 Velocity Components at a Blade Section

Consider a blade element at a distance $r$ from the rotor hub shown in Figure 9. For ease of analysis (mainly for 2-D flow assumptions to hold), we consider only the planar component of velocity i.e. $|V_h| = |V_x, V_y, 0|$, $|v_h^i| = |v_x^i, v_y^i, 0|$ and vertical velocity $V_z$. To reduce notational confusion, the $|.|$ around the induced horizontal velocity and horizontal induced velocity will be dropped. The total airflow velocity at the blade element is $U(r, \psi) \in \mathbb{R}^3$ and $U(r, \psi) \in \{D\}$. The transverse scalar velocity at a blade element $U_\mathrm{h}(r, \psi) \in \mathbb{R}$ which is the magnitude of the planar projection of $U(r, \psi)$ in $\vec{e}_1, \vec{e}_2 \in \{D\}$ is

$$U_\mathrm{h}(r, \psi) = \varpi r + (V_\mathrm{h} + v_h^\mathrm{i})\sin\psi. \tag{31}$$

For the vertical velocity $U_\mathrm{z}(r, \psi)$

$$U_\mathrm{z}(r, \psi) = v_\mathrm{z}^\mathrm{i} - V_z + r\dot{\beta}(\psi) + (v_h^i + V_h)\beta(\psi)\cos\psi. \tag{32}$$

Normalising or non-dimensionalising by dividing by the tip velocity of the rotor $\varpi R$, the following are obtained

$$u_z(r,\psi) = \frac{U_z(r,\psi)}{\varpi R} = \lambda + \frac{r}{R\varpi}\dot\beta(\psi) + \frac{1}{R\varpi}(v_h^i + V_h)\beta(\psi)\cos\psi,$$
$$= \lambda + \frac{r}{R}\frac{\mathrm{d}\beta(\psi)}{\mathrm{d}\psi} + \mu\beta(\psi)\cos\psi, \tag{33}$$

and

$$u_h(r,\psi) = \frac{U_h(r,\psi)}{\varpi R},$$
$$= \frac{r}{R} + \mu\sin\psi. \tag{34}$$

The total or resultant velocity at the blade element is

$$|U(r,\psi)| = \sqrt{U_h(r,\psi)^2 + U_z^2(r,\psi)}.$$

**Assumption 3.2** *We assume that $U^2(r,\psi) \cong U_h^2(r,\psi)$ for the quadrotor used in this report.*

This is a valid assumption as the quadrotor which weighs under 2kg, experiments have shown that for its rotor of radius $10in$ with $\varpi \geq 5000 RPM$, the velocity of the vehicle is bounded and in this case $|V| \leq 5m/s$. This implies that if the vehicle is doing an axial motion at this maximum velocity such that $U_h(r,\psi) = 52.3m/s$ and $U_z(r,\psi) = 5m/s$. Hence $U_h(r,\psi) > 10 U_z(r,\psi)$ from which it is easily seen that $U^2(r,\psi) \cong U_h^2(r,\psi)$.

### 3.6 Aerodynamic Forces acting on Blade Elements

The aerodynamic forces acting on a blade element shown in Figure 9 are defined as

**Lift** $\mathrm{d}L(r,\psi) \in \mathbb{R}$ is the force generated by the blade element perpendicular to the direction of the resultant airflow $U(r,\psi)$.

**Drag** $\mathrm{d}D(r,\psi) \in \mathbb{R}$ is the force generated by the blade element that is parallel to the direction of the resultant airflow.

The elemental lift and drag forces on a blade element expressed in the TPP $\{D\}$ are defined by

$$dL(r,\psi) = \frac{1}{2}\rho U(r,\psi)^2 C_l(r,\psi) c(r)\,\mathrm{d}r, \tag{35}$$
$$dD(r,\psi) = \frac{1}{2}\rho U(r,\psi)^2 C_d(r,\psi) c(r)\,\mathrm{d}r, \tag{36}$$

where $C_l(r,\psi)$ is the lift coefficient, $C_d(r,\psi)$ the drag coefficient and $c(r)$ is the chord length at a section radius $r$ from the hub. The coefficients $C_l(r,\psi)$ and $C_d(r,\psi)$ are expressed respectively as

$$C_l(r,\psi) = C_{l_0} + C_{l\alpha}\alpha(r,\psi), \tag{37}$$
$$C_d(r,\psi) = C_{d_0} + K C_l(r,\psi)^2, K > 0. \tag{38}$$

For a $3-D$ wing planform, the constant $K = \frac{1}{\pi ARe}$, where $AR$ is the *aspect ratio* of the wing and $e$ is the Oswald span efficiency. The $AR$ for helicopter blades is usually large $> 10$ and $e = 0.8$ for an elliptical lift distribution. From Figure 9, the blade element angle of attack $\alpha(r,\psi)$ which is the angle between the mean chord line of the aerofoil and the direction of motion of the blade or airflow is defined as

$$\alpha(r,\psi) = \theta(r) - \phi(r,\psi), \tag{39}$$

where $\theta(r)$ is the blade pitch and $\phi(r,\psi)$ is the relative inflow angle at the blade section defined by

$$\phi(r,\psi) = \tan^{-1}\frac{U_z(r,\psi)}{U_h(r,\psi)}, \tag{40}$$

for which we make the following assumption.

**Assumption 3.3** *The relative inflow angle $|\phi(r,\psi)| \leq 10°$ is small such that $\cos\phi(r,\psi) \cong 1$ and $\sin\phi(r,\psi) \cong \phi(r,\psi)$ for all $r$ and $\psi$.*

With this assumption, (40) becomes
$$\phi(r,\psi) \cong \frac{U_z(r,\psi)}{U_h(r,\psi)}. \tag{41}$$

Hence (39) can be approximated by
$$\alpha(r,\psi) \cong \theta(r) - \frac{U_z(r,\psi)}{U_h(r,\psi)}. \tag{42}$$

**Remark 7** *The most ideal or optimum rotor geometry is one that maintains a constant angle of attack and constant induced velocity across the entirety of the blade [6, pg. 54]. To produce a constant angle of attack for purely axial flights, the optimum rotor makes use of a hyperbolic pitch geometry from the hub to the tip of the rotor. In addition, using a hyperbolic geometry for the chord ensures constant spanwise induced velocity. In [15, 16], the authors designed a slight variation of such an optimum rotor for the ANU-X4 flyer. It should be noted that there are physical limitations on the rotor around the hub such that $\lim_{r \to 0} c(r), \theta(r) \to \mathfrak{R}^+$ or $c(r), \theta(r)$ have physical values. In addition the 20 to 30% rotor around the hub is curved inwards to prevent large horizontal forces, torque and practicallity of manufacture. The blade element chord and pitch at a distance $r$ from the hub for the ideal/optimum rotor are defined by the following hyperbolic functions*

$$c(r) = \frac{c_{tip}}{r/R},$$

$$\theta(r) = \frac{\theta_{tip}}{r/R},$$

*where $c_{tip}$ ad $\theta_{tip}$ are the tip chord and pitch respectively, $r \in [0,R]$ is the distance from the rotor hub and $R$ is the rotor radius.*

The theoretical development of models for $T, H$ and $P$ using the ideal blade geometry with aerofoils for which along the span and every azimuth angle $\psi$, $C_l(r,\psi)$ and $C_d(r,\psi)$ are defined by (37) and (38) respectively are carried out in Section 6.

From the elemental forces defined by (35) and (36) are the forces along the $\vec{e}_1, \vec{e}_2$ plane and the $\vec{e}_3$ of the tip path plane $\{D\}$. These horizontal and vertical forces defined in $\{D\}$ are given by

$$dF_x(r,\psi) = dL(r,\psi)\sin\phi(r,\psi) + dD(r,\psi)\cos\phi(r,\psi), \tag{43}$$
$$dF_z(r,\psi) = dL(r,\psi)\cos\phi(r,\psi) - dD(r,\psi)\sin\phi(r,\psi). \tag{44}$$

It should be noted that $F_x$ represents the magnitude of the force along the $x-y$ plane or plane containing $\vec{e}_1, \vec{e}_2$. Furthermore we define the following elemental forces along the rotor hub or body fixed frame $\{B\}$.

**In-plane H force** $dH(r,\psi) \in \mathbb{R}$ is the resultant elemental force generated on the plane of $\vec{e}_1, \vec{e}_2$ of $\{B\}$ that opposes the motion along the $\vec{e}_1, \vec{e}_2$ plane of $\{B\}$.

**Thrust** $dT(r,\psi) \in \mathbb{R}$ is the resultant elemental force generated along $\vec{e}_3$ of $\{B\}$.

In the next sections, derivations of these elemental forces and associated elemental torque and their respective sums in $\{B\}$ will be presented in accordance with Figure 10 which shows the rotor reference frame ($\{C\}$) with the horizontal and vertical forces that are tilted from $\{B\}$ by the flapping angle $\beta(\psi)$. The derivations are carried out in the body fixed frame of the rotor ($\{B\}$) or the rotor hub or shaft for different blade geometries and aerodynamic characteristics of the aerofoil section and rotor aspect ratio $AR$.

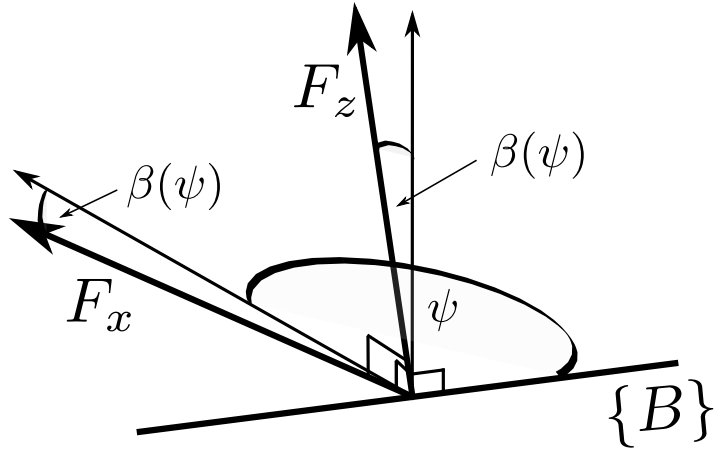

Figure 10: Blade element forces in $\{B\}$ and $\{D\}$. The figure shows the horizontal and vertical forces in the TPP $\{D\}$ rotated by $\beta(\psi)$ from the rotor hub/body fixed frame $\{B\}$.

## 3.7 Induced Power Factor $\kappa$

The *induced power factor* $\kappa$ is a factor that accounts for the additional power/energy dissipated due to wake rotation, tip loss and non-uniform flow that is not modelled by momentum theory. These power lost effects are more significant for small rotors such as quadrotor blades with high disc loading and low power efficiency $\left(\frac{T}{P}\right)$ that are generally less efficient than helicopters. It changes with changing aerodynamic conditions around a rotor and increases with increasing tip loss and decreases with increasing rotor efficiency. These aerodynamic losses only apply to the induced power component of power. The *induced power factor* is closely related to but not the same as the *figure of merit*. Unlike the *figure of merit* used in the analysis of full scale helicopters, $\kappa$ does not model profile power losses. This $\kappa$ contains $\kappa_e$ described in Section 2.5 which is incorporated to account for the use of the assumption of uniform and constant induced inflow velocities.

The *induced power factor* can be better explained in terms of disc loading $DL$. For helicopters with low disc loading i.e. large rotor areas relative to the thrust they produce, they have a high thrust coefficient $C_T$ and high power efficiency $\left(\frac{T}{P}\right)$ than quadrotors. Furthermore, the following also apply for a helicopter

1. Given the disc area and high torque engines for helicopter rotors implies that they require less RPM compared to quadrotors. For example the blades on the quadrotor under study require $RPM \approx 5000$ to hover while those on normal helicopters require significantly less.

2. The induced velocity $v^i$, is lower for helicopters compared to quadrotors at hover. Hence, to maintain the low angle of attack, helicopter blades have low collective pitch at hover. The higher $C_T$ is as a result of higher rotor radius as $C_T$ is proportional to the cube of the radius.

3. The high rotor efficiency at hover is as a result of the high thrust generated by large area with small induced velocity where in hover the total airflow through the rotor is $v_z^i$ and therefore low required power.

4. The low disc loading on helicopter blades implies that they are under less back pressure than quadrotor blades.

5. Because the thrust is very high with less back pressure, axial relative wind only slightly affects the back pressure and the thrust produced at a given power. It should be noted that the thrust and power are very high thus there is no significant change in the efficiency i.e. $\frac{T}{P}$ of the rotor with axial wind. Note however that given the low $RPM$ of the rotors implies that small changes in thrust results in observable changes in $C_T$.

So in the presence of an updraft, there is only a slight increase in rotor efficiency despite an increase in $C_T$. The dominant effect is the additional work done by the rotors as a result of increased swirl in the wake as well as additional tip losses in the generation of tip vortices. Thus increasing $C_T$ as a result of an updraft causes additional losses with little changes in rotor efficiency and therefore corresponds to a moderate increase in $\kappa$.

For high disc loading rotors with low thrust coefficients $C_T$ such as quadrotors,

1. The angle of attack of such rotors is very high which corresponds to higher blade pitch angle in static free air or at hover. It should be noted that in such aerodynamic condition, given the small rotor disc area implies that to produce thrust requires higher induced velocity $v_z^i$ which corresponds to higher power and therefore lower efficiency than do helicopter rotors.

2. With an updraft, the thrust produced increases, hence $C_T$ and a decrease in total velocity through the rotor. This leads to an increase in rotor efficiency. The relative high disc loading, low thrust and power requirements including high $\varpi$ compared to helicopters implies that adding or removing power into the system will have a significant effect on the rotor efficiency. This can be illustrated mathematically by using the rotor efficiency equations (for e.g. (15)) and recognising the low profile power requirement for quadrotor blades compared to helicopter blades.

Hence for quadrotor rotor blades an updraft increases $C_T$ slightly (due to high $\varpi$), increases efficiency significantly due to low thrust and power and a negligible change in the already high tip loss. Hence overall for high disc loading low $C_T$ quadrotor blades, $\kappa$ decreases with increasing $C_T$.

With these intuitions and with reference to Figure 11, we propose the following general model relating $\kappa$ to $C_T$

$$\kappa = d_0 + d_1 \frac{1}{C_T} + d_2 C_T. \tag{45}$$

It should be noted that the model illustrated in Figure 11 is supported by Figure 3.18 [9, pg. 105] although Leishman only considers low disc loading helicopter rotor blades. In the region of operation of quadrotors ($C_T << 10^{-4}$), the dominant part of the model is $d_1 \frac{1}{C_T}$. For helicopters with large $C_T$, the dominant part is $d_2 C_T$.

Given that the dominant part of the *induced power factor* model (45) for quadrotors ($C_T << 10^{-4}$) is $d_1 \frac{1}{C_T}$ and if $\bar{C}_T$ is an operating point of the rotor, then it can be shown algebraically that

$$d_1 \frac{1}{C_T} = \frac{const_1}{\bar{C}_T} - \frac{const_2}{\bar{C}_T^2} \left( C_T - \bar{C}_T \right),$$

where the constants *const* are some arbitrary constants. Hence one can approximate the $\kappa$ model for quadrotors by a linear model

$$\kappa = \beta_0 + \beta_1 C_T, \tag{46}$$

where $\beta_0 > 0$ and $\beta_1 < 0$ is a large negative constant. With this linear model, there is a significant reduction in the computational requirement for $\kappa$ when implementing on computationally constrained embedded *electronic speed controllers* used on quadrotors. In the derivations carried out in the sequel, the induced velocity components of the power contains $\kappa$. To reduce the number of variables during the development of the models, we make the following remark.

**Remark 8** *To reduce the many variables in the torque/power derivations, we will use $\lambda$ in the power models and later on account for the effects mentioned in the final model by replacing it with $\kappa \lambda^i + \lambda_z$ or $\kappa v_z^i - V_z$. This also applies to the contribution of the horizontal force i.e. $\kappa \mu^i + \mu_h$ or $\kappa v_h^i + V_z$.*

## 4 Blade Element Theory for Classical Rotor Geometry (Constant chord and pitch) and Infinite Aspect Ratio

In this section, we apply the elemental forces obtained in Section 3 to model $T, H$ and $\tau$ hence power $P$ of the entire rotor blade in $\{B\}$. The blade geometry used in the analysis is the simplest geometry which consists of a constant chord $c(r) = c$, constant pitch $\theta(r) = \theta_0$ and a blade of infinite aspect ratio (AR) of length $R$. In addition, the rotor aerofoil considered has zero-lift angle of attack i.e. $C_{l0} = 0$ and a linear lift slope. This is the rotor geometry and aerofoil characteristics considered in the analysis contained in the helicopter literature [6, 9, 21]. Though this rotor is far from the rotors used on quadrotors, it is however a good starting point for modelling. The final models obtained are simplified to obtain lumped parameter models that can be used for Robotic applications.

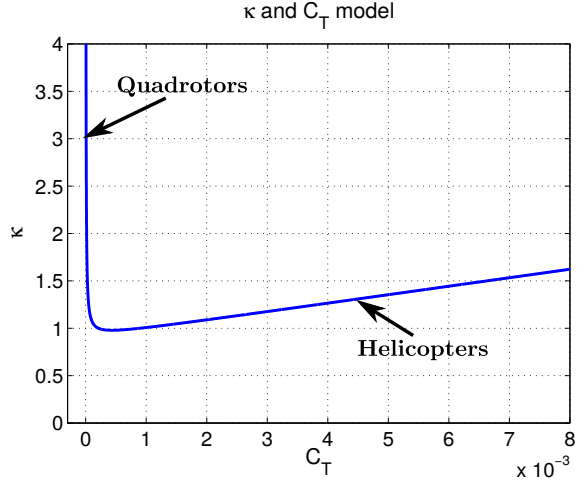

Figure 11: An illustration of the induced power factor $\kappa$ and thrust coefficient $C_T$ for low and high disc loading rotor blades used on quadrotors and helicopters respectively.

With $AR = \infty$ implies $K = 0$ and $C_{l0} = 0$, the lift coefficient (37) and drag coefficient (38) become

$$C_l(r, \psi) = C_{l\alpha}\alpha(r, \psi),$$
$$C_d(r, \psi) = C_{d0} = C_d.$$

Therefore from (42)

$$C_l(r, \psi) \cong C_{l\alpha}\left(\theta(r) - \frac{U_z(r, \psi)}{U_h(r, \psi)}\right), \tag{47}$$

which implies that with Assumption 3.2, (35) becomes

$$dL(r, \psi) \cong \frac{1}{2}\rho U_h(r, \psi)^2 C_{l\alpha}\left(\theta(r) - \frac{U_z(r, \psi)}{U_h(r, \psi)}\right)c(r)\,dr, \tag{48}$$

and the elemental drag force (36)

$$dD(r, \psi) \cong \frac{1}{2}\rho U_h(r, \psi)^2 C_d c(r)\,dr. \tag{49}$$

### 4.1 Rotor Thrust

Classical helicopter theory for this is covered in [6, pg. 96-98] and [21, pg. 58-60]. From the definitions given in Section 3.6, an expression for the thrust modulus $dT(r, \psi)$ or the elemental thrust can be obtained. To do this, consider again Figure 10 and resolving forces in the $\vec{e}_3$ direction of $\{B\}$ for a blade element,

$$dT(r, \psi) = dF_z(r, \psi)\cos\beta(\psi) + dF_x(r, \psi)\sin\beta(\psi).$$

Substituting for $dF_x$ and $dF_z$,

$$dT(r, \psi) = [dL(r, \psi)\cos\phi(r, \psi) + dD(r, \psi)\sin\phi(r, \psi)]\cos\beta(\psi)$$
$$+ [dL(r, \psi)\sin\phi(r, \psi) + dD(r, \psi)\cos\phi(r, \psi)]\sin\beta(\psi).$$

Realising that $dD(r,\psi)\sin\phi(r,\psi)\cos\beta(\psi)$ and $dD(r,\psi)\cos\phi(r,\psi)\sin\beta(\psi)$ both consist of two small terms that can be neglected i.e.

$$dD(r,\psi)\sin\phi(r,\psi)\cos\beta(\psi) \cong 0,$$
$$dD(r,\psi)\cos\phi(r,\psi)\sin\beta(\psi) \cong 0.$$

Hence,
$$dT(r,\psi) = dL(r,\psi)\cos\phi(r,\psi)\cos\beta(\psi) + dL(r,\psi)\sin\phi(r,\psi)\sin\beta(\psi).$$

In addition, $dL(r,\psi)\sin\phi(r,\psi)\sin\beta(\psi)$ consists of two small terms hence its effect is also negligible i.e.

$$dL(r,\psi)\sin\phi(r,\psi)\sin\beta(\psi) \cong 0.$$

Therefore the elemental thrust is approximated by

$$dT(r,\psi) \cong dL(r,\psi). \tag{50}$$

Hence,
$$\begin{aligned}dT(r,\psi) &\cong \frac{1}{2}\rho U_h(r)^2 C_{l\alpha}\left(\theta(r)-\frac{U_z(r,\psi)}{U_h(r,\psi)}\right)c(r)dr, \\ &= \frac{1}{2}\rho C_{l\alpha}\left(\theta(r)U_h(r,\psi)^2 - U_z(r,\psi)U_h(r,\psi)\right)c(r)\mathrm{d}r.\end{aligned} \tag{51}$$

Substituting for $U_z(r,\psi)$ and $U_h(r,\psi)$ using their normalised forms (33) and (34) and from (16), (51) becomes

$$T = \frac{N_b}{2\pi}\int_0^{2\pi}\int_0^R \frac{1}{2}\rho C_{l\alpha}\varpi^2 R^2\left[\underbrace{\theta(r)\left(\frac{r}{R}+\mu\sin\psi\right)^2}_{T_\theta} - \underbrace{u_z\left(\frac{r}{R}+\mu\sin\psi\right)}_{T_\phi}\right]c(r)\mathrm{d}r\mathrm{d}\psi,$$

Knowing that $\int_0^{2\pi}\sin\psi = 0$ along with all other odd powers of $\sin\psi$,

$$T_\theta = \frac{N_b}{2\pi}\int_0^{2\pi}\int_0^R \frac{1}{2}\rho C_{l\alpha}R^2\varpi^2\left[\theta(r)\left(\left(\frac{r}{R}\right)^2+\mu^2\sin^2\psi\right)\right]c(r)\mathrm{d}r\mathrm{d}\psi.$$

Given that we are using a constant chord $c(r)=c$ and pitch $\theta(r)=\theta_0$,

$$\begin{aligned}T_\theta &= \frac{N_b}{2\pi}\int_0^{2\pi}\int_0^R \frac{1}{2}\rho c C_{l\alpha}\varpi^2 R^2\left[\theta_0\left(\left(\frac{r}{R}\right)^2+2\left(\frac{r}{R}\right)\mu\sin\psi+\mu^2\sin^2\psi,\right)\right]\mathrm{d}r\mathrm{d}\psi \\ &= \frac{1}{4}N_b\rho c C_{l\alpha}cR^3\varpi^2\left[\frac{2}{3}\theta_0\left(1+\frac{3}{2}\mu^2\right)\right].\end{aligned}$$

Consider now the $\phi$ component of $T$ and substituting for $\beta(\psi)$ and $\frac{\mathrm{d}\beta(\psi)}{\mathrm{d}\psi}$

$$\begin{aligned}dT(r,\psi)_\phi &= \frac{1}{2}\rho c C_{l\alpha}\varpi^2 R^2\left(\lambda+\frac{r}{R}\frac{\mathrm{d}\beta(\psi)}{\mathrm{d}\psi}+\mu\beta(\psi)\cos\psi\right)\left(\frac{r}{R}+\mu\sin\psi\right)c\mathrm{d}r\mathrm{d}\psi, \\ &= \frac{1}{2}\rho c C_{l\alpha}\varpi^2 R^2\left(\lambda+\frac{r}{R}(a_1\sin\psi-b_1\cos\psi)+\mu(a_0-a_1\cos\psi-b_1\sin\psi)\cos\psi\right)\left(\frac{r}{R}+\mu\sin\psi\right)c\mathrm{d}r\mathrm{d}\psi, \\ dT(r)_\phi &= \frac{1}{2}N_b\rho c C_{l\alpha}\varpi^2 R^2\left(\lambda\frac{r}{R}+\frac{r}{2R}a_1\mu-a_1\mu\frac{r}{2R}\right)c\mathrm{d}r, \\ T_\phi &= \frac{1}{4}N_b\rho c C_{l\alpha}\varpi^2 R^3\lambda\end{aligned}$$

Hence using blade element momentum theory, the thrust $T$ is modelled by

$$T = T_\theta - T_\phi,$$
$$T = \frac{1}{4}N_b \rho c C_{l\alpha} c R^3 \varpi^2 \left[\frac{2}{3}\theta_0\left(1 + \frac{3}{2}\mu^2\right) - \lambda\right]. \tag{52}$$

To illustrate the effect of nonuniform and non-constant distribution of $v^i$, consider the induced velocity model for $v^i_z$ from Section 3.4

$$v^i_z(r, \psi) = 4v^{i0}_z \left(\frac{1}{2}d_0\right),$$

then the $v^i_z$ component of the thrust which is embedded in $\lambda$ based on (51) is given by

$$dT(r, \psi)_{vi} = \frac{N_b}{2\pi} \int_0^R \int_0^{2\pi} \left(\frac{1}{2}\rho c C_{l\alpha} R \varpi\right) 4d_0 \left(\frac{1}{2}v^{i0}_z\right)\left(\frac{r}{R} + \mu \sin\psi\right) d\psi dr,$$

$$= \frac{15}{8}\rho c N_b C_{l\alpha} R \varpi \int_0^R \sqrt{1 - \left(\frac{r}{R}\right)^2} \left(\frac{r}{R}\right)^3 v^{i0}_z dr,$$

$$T_{vi} = \frac{1}{4}\rho c N_b C_{l\alpha} R^2 \varpi v^{i0}_z.$$

This shows that there is no change in the equation obtained for $T$ in (52) since $\lambda = \frac{v^i_z - V_z}{\varpi R}$ and therefore reaffirms our use of $v^i(r, \psi) = v^i = v^{i0}$ in the computation of $T$. If any effects, non-constant $v^i$ will affect the $P$ model as outined in Remark 8.

## 4.2 Rotor In-Plane H-Force Component

Classical helicopter theory for this is covered in [6, pg. 98-100] and [21, pg. 60-61]. It is now left to show derivation of the drag force opposing the motion of the rotor plane which is otherwise known as the the rotor in-plane horizontal force $H$. Referring again to Figure 10, the force on the horizontal plane of $\{B\}$ on the rotor hub is resolved as

$$dH(r, \psi) = -\left(dF_x(r, \psi)\cos\beta(\psi)\sin\psi + dF_z(r, \psi)\sin\beta(\psi)\cos\psi\right).$$

It should be noted that with the vehicle moving forward i.e. $V_h > 0$ shown in Figure 9, this force is pointed in the negative direction to the direction of motion. Hence it is negative and therefore opposes the direction of motion. However, the analysis will be carried out to determine the magnitude of the force. Given that $\beta(\psi)$ is small and substituting for $dF_x$ and $dF_z$ using (43) and (44) respectively, the following is obtained

$$dH(r, \psi) = \underbrace{dD(r, \psi) \sin\psi}_{dH_\text{P}(r,\psi)} + \underbrace{dL(r, \psi)\left(\beta(\psi)\cos\psi + \phi(r, \psi)\sin\psi\right)}_{dH_\text{i}(r,\psi)}, \tag{53}$$

where $dH_i(r, \psi)$ is known as the induced in-plane force as it is induced by the inclination of the elemental lift vector $dL(r, \psi)$ and $dH_P(r, \psi)$ is the profile drag as it is the contribution of the elemental drag $dD(r, \psi)$ and it is the drag force generated by the transverse velocity of the rotor blades through the air [4]. The profile ($H_P$) contribution to $H$ is obtained as follows

$$H_\text{p} = \frac{N_b}{2\pi} \int_0^{2\pi} \int_0^R dH_\text{P}(r, \psi) dr d\psi = \frac{N_b}{2\pi} \int_0^{2\pi} \int_0^R dD(r) \sin\psi dr d\psi.$$

Substituting for $dD(r, \psi)$ using (36)

$$H_\text{p} = \frac{N_b}{2\pi} \int_0^{2\pi} \int_0^R \frac{1}{2}\rho U(r, \psi)^2 C_d(r, \psi) c(r) \sin\psi dr d\psi. \tag{54}$$

From Assumption 3.2, $U^2(r,\psi) \cong U_h^2(r,\psi)$ and substituting for $U_h(r,\psi) = \mu\varpi R$ from the definitions,

$$H_{\rm p} = \frac{N_b}{2\pi} \int_0^{2\pi} \int_0^R \frac{1}{2}\rho\varpi^2 R^2 \left(\left(\frac{r}{R}\right)^2 + 2\left(\frac{r}{R}\right)\mu\sin\psi + \mu^2\sin^2\psi\right) C_d(r,\psi) c(r) \sin\psi \, dr \, d\psi. \tag{55}$$

Once more, it should be noted that we are using a constant induced velocity such that $v^i(r,\psi) = v^i$ and therefore there is no need to consider the Mangler and Squire expressions for the induced velocities presented in Section 3.4. Using $c(r) = c, \theta(r) = \theta_0, C_d(r,\psi) = C_d, C_l(r,\psi) = C_{l\alpha}\alpha(r)$, the integral in (55) becomes

$$H_{\rm p} = \frac{1}{4}\rho N_b c C_d R^3 \mu \varpi^2. \tag{56}$$

This is the in-plane horizontal force or drag force generated as a result of forward motion of the rotor disc that does not involve the contribution from the lift force. The lift induced component $dH_i(r,\psi)$ is obtained by substituting for $dL(r,\psi)$ using (48) and substituting for and using the small angle assumption for $\phi(r,\psi)$

$$dH_i(r,\psi) = dL(r,\psi)\left(\beta(\psi)\cos\psi + \phi(r,\psi)\sin\psi\right),$$
$$= \frac{1}{2}\rho U^2(r,\psi)C_{l\alpha}\left(\theta_0 - \frac{U_z(r,\psi)}{U_h(r,\psi)}\right)\left[\beta(\psi)\cos\psi + \frac{U_z(r,\psi)}{U_h(r,\psi)}\sin\psi\right] c \, dr.$$

Expanding the equation with $U^2(r,\psi) \cong U_h(r,\psi)^2$ (Assumption 3.2),

$$dH_i(r,\psi) = \frac{1}{2}\rho C_{l\alpha}\left[\left(\theta_0 U_h(r,\psi)^2 - U_z(r,\psi)U_h(r)\right)\beta(\psi)\cos\psi + \left(\theta_0 U_z(r,\psi)U_h(r,\psi) - U_z(r,\psi)^2\right)\sin\psi\right] c \, dr. \tag{57}$$

The overall induced drag $H_i$ is given by

$$H_i = \frac{N_b}{2\pi}\int_0^{2\pi}\int_0^R dH_i(r,\psi) \, d\psi,$$

$$= \frac{N_b}{2\pi}\int_0^{2\pi}\int_0^R \frac{1}{2}\rho c C_{l\alpha}\left[\underbrace{\left(\theta_0 U_h^2(r,\psi) - U_h(r,\psi)U_z(r,\psi)\right)\beta(\psi)\cos\psi}_{H_{i\beta}} + \underbrace{\left(\theta_0 U_z(r,\psi)U_h(r) - U_z(r,\psi)^2\right)\sin\psi}_{H_{i\phi}}\right] dr d\psi.$$

Taking first the $\phi$ component,

$$H_{i\phi} = \frac{N_b}{2\pi}\int_0^{2\pi}\int_0^R \frac{1}{2}\rho c C_{l\alpha}\left(\theta_0 U_h(r,\psi)U_z(r,\psi) - U_z^2(r,\psi)\right)\sin\psi \, dr d\psi,$$

$$= \frac{N_b}{2\pi}\int_0^{2\pi}\int_0^R \frac{1}{2}\rho c C_{l\alpha} R^2 \varpi^2\left(\theta_0\left(\frac{r}{R} + \mu\sin\psi\right) u_z - u_z^2\right)\sin\psi \, dr d\psi,$$

$$H_{i\phi} = \frac{N_b}{2\pi}\int_0^{2\pi}\int_0^R \frac{1}{2}\rho c C_{l\alpha}\theta_0 R^2 \varpi^2\left(\theta_0\left(\frac{r}{R} + \mu\sin\psi\right)\left(\lambda + \frac{r}{R}\frac{d\beta(\psi)}{d\psi} + \mu\beta(\psi)\cos\psi\right) - \left(\lambda + \frac{r}{R}\frac{d\beta(\psi)}{d\psi} + \mu\beta(\psi)\cos\psi\right)^2\right)\sin\psi \, dr d\psi.$$

Therefore,

$$H_{i\phi} = \frac{1}{4}N_b\rho c C_{l\alpha} R^3 \varpi^2\left(\theta_0\left(\frac{a_1}{3} - \frac{a_1\mu^2}{4} + \lambda\mu\right) - a_1\lambda + \frac{a_1^2\mu}{4} - \frac{b_1^2\mu}{4} + \frac{a_0 b_1\mu^2}{2}\right).$$

However, the terms with products of the flapping coefficients and products with $\mu^2$ given that they are very small, results in

$$H_{i\phi} = \frac{1}{4} N_b \rho c C_{l\alpha} R^3 \varpi^2 \left( \theta_0 \left( \lambda\mu + \frac{1}{3} a_1 \right) - \lambda a_1 \right).$$

Then the $H_{i\beta}$ component is

$$H_{i\beta} = \frac{N_b}{2\pi} \int_0^{2\pi} \int_0^R \frac{1}{2} \rho c C_{l\alpha} U_h^2(r,\psi) \left( \theta_0 - \frac{U_z(r,\psi)}{U_h(r,\psi)} \right) \beta(\psi) \cos\psi \, \mathrm{d}r \mathrm{d}\psi,$$

$$= \frac{N_b}{2\pi} \int_0^{2\pi} \int_0^R \frac{1}{2} \rho c C_{l\alpha} \left( \theta_0 U_h^2(r,\psi) - U_z(r,\psi) U_h(r,\psi) \right) \beta(\psi) \cos\psi \, \mathrm{d}r \mathrm{d}\psi.$$

Hence,

$$H_{i\beta} = \frac{N_b}{2\pi} \int_0^{2\pi} \int_0^R \frac{1}{2} \rho c C_{l\alpha} \left( \theta_0 \left( \frac{r}{R} + \mu\sin\psi \right)^2 - \left( \lambda + \frac{r}{R} \frac{\mathrm{d}\beta(\psi)}{\mathrm{d}\psi} + \mu\beta(\psi)\cos\psi \right) \left( \frac{r}{R} + \mu\sin\psi \right) \right) (a_0 - a_1\cos\psi - b_1\sin\psi) \cos\psi \, \mathrm{d}r \mathrm{d}\psi.$$

Therefore

$$H_{i\beta} = \frac{1}{4} \rho C_{l\alpha} N_b c R^3 \varpi^2 \left[ a_0 b_1/3 + a_1 \lambda/2 - a_1 \theta_0/3 - a_0^2 \mu/2 - a_1^2 \mu/4 - b_1^2 \mu/4 - a_1 \mu^2 \theta_0/4 + a_0 b_1 \mu^2/2 \right].$$

Given that the flapping coefficients are small for quadrotor blades implies that their products are negligible. Hence

$$H_{i\beta} = \frac{1}{4} \rho C_{l\alpha} N_b c R^3 \varpi^2 \left[ \frac{a_1 \lambda}{2} - \frac{a_1 \theta_0}{3} \right].$$

Hence the overall induced H-force is

$$H_\mathrm{i} = H_{i\phi} + H_{i\beta},$$

$$H_\mathrm{i} = \frac{1}{4} N_b \rho C_{l\alpha} c R^3 \varpi^2 \left[ \theta_0 \left( \lambda\mu + \frac{1}{3} a_1 \right) - \lambda a_1 + \frac{a_1 \lambda}{2} - \frac{a_1 \theta_0}{3} \right],$$

$$H_\mathrm{i} = \frac{1}{4} N_b \rho C_{l\alpha} c R^3 \varpi^2 \mu \left[ \theta_0 \left( \lambda + \frac{2}{3}(4\theta_0/3 - \lambda) \right) - \lambda(4\theta_0/3 - \lambda) - \frac{2}{3} \theta_0(4\theta_0/3 - \lambda) \right]. \tag{58}$$

For very rigid blades the flapping coefficients $a_0, a_1, b_1$ are very small resulting in a very small $H_i$. This is the reason quadrotor blades are designed to be very rigid.

This is the additional component of the $H$-force as a result of the tilting of the tip path plane with respect to the hub $\{B\}$. Therefore, the total in-plane horizontal force $H$ is given by

$$H = H_p + H_i,$$

$$H = \frac{1}{4} N_b \rho C_{l\alpha} c R^3 \varpi^2 \mu \left[ \frac{C_d}{C_{l\alpha}} + \theta_0 \left( \lambda + \frac{2}{3}(4\theta_0/3 - \lambda) \right) - \lambda(4\theta_0/3 - \lambda) - \frac{2}{3} \theta_0(4\theta_0/3 - \lambda) \right],$$

By making use of the fact that $\theta_0$ is small such that $\theta_0^2 \approx 0$,

$$H = \frac{1}{4} N_b \rho C_{l\alpha} c R^3 \varpi^2 \mu \left[ \frac{C_d}{C_{l\alpha}} - \frac{1}{3} \theta_0 \lambda + \lambda^2 \right] \tag{59}$$

For low pitched blades, the $H$-force decreases with increasing $\lambda$. Hence $H$-force is directly linked to $T$. In the current quadrotor literature, a linear relation $H \propto T$ has been proposed in [2, 13].

## 4.3 Aerodynamic Torque ($\tau$) and Power ($P$)

This model is covered in [6, pg. 100-103] and [21, pg. 61-62] for helicopters. In order to determine the rotational torque created around the rotor hub or $\vec{e}_3$ of $\{B\}$, consider again Figure 9 from which the elemental torque created by a blade element at a distance $r$ from the hub is given by

$$d\tau(r,\psi) = r\left(dD(r,\psi)\cos\phi(r,\psi) + dL(r,\psi)\cos\beta(\psi)\sin\phi(r,\psi)\right)drd\psi,$$
$$= \underbrace{rdD(r,\psi)}_{d\tau_\text{P}} + \underbrace{rdL(r,\psi)\phi(r,\psi)}_{d\tau_i(r,\psi)} drd\psi, \tag{60}$$

where $d\tau_\text{P}$ is the torque required for spinning the rotor and is termed profile drag and $d\tau_i$ is the induced torque as a result of the tilting of the thrust force. The profile drag torque $\tau_\text{P}$ is given by

$$\tau_\text{P} = \frac{N_b}{2\pi}\int_0^{2\pi}\int_0^R d\tau_\text{P}(r,\psi)\,d\psi = \frac{N_b}{2\pi}\int_0^{2\pi}\int_0^R rdD(r,\psi)drd\psi.$$

Substituting for $dD(r,\psi)$ using (36)

$$\tau_\text{P} = \frac{N_b}{2\pi}\int_0^{2\pi}\int_0^R \frac{1}{2}\rho U(r,\psi)^2 C_d(r,\psi)c(r)r\,dr\,d\psi.$$

Substituting for $U^2(r,\psi)$ with $U^2(r,\psi) \cong U_h^2(r,\psi)$ (Assumption 3.2), $c(r) = c$ and $C_d(r,\psi) = C_d$,

$$\tau_\text{P} = \frac{N_b}{2\pi}\int_0^{2\pi}\int_0^R \frac{1}{2}\rho c C_d \varpi^2 R^2 \left(\frac{r}{R} + \mu\sin\psi\right)^2 r\,dr\,d\psi,$$
$$= \frac{1}{8}\rho N_b c C_d \varpi^2 R^4 \left(1+\mu^2\right). \tag{61}$$

For the induced torque ($\tau_i$) component of (60)

$$d\tau_i(r,\psi) = \frac{1}{2}\rho c U^2(r,\psi)C_{l\alpha}\alpha\phi(r,\psi)rdrd\psi,$$
$$= \frac{1}{2}\rho c C_{l\alpha}(\theta_0 U_z(r,\psi)U_h(r,\psi) - U_z^2(r,\psi))rdrd\psi.$$

Torque hence power as pointed out in Section 2 always has components for the generation of $T$ and $H$. In order to get $\tau$ in terms of these forces, we first substitute for $r$ from the definition of $U_h(r,\psi)$ given by (31) i.e. $r = \frac{U_h(r,\psi) - (v_h^i + V_h)\sin\psi}{\varpi}$. So that

$$d\tau_i(r,\psi) = \frac{1}{2}\frac{\rho C_{l\alpha}c}{\varpi}(\theta_0 U_z(r,\psi)U_h(r,\psi) - U_z^2(r,\psi))(U_h(r,\psi) - (v_h^i + V_h)\sin\psi)drd\psi.$$

Expanding this and substituting for $dT(r,\psi)$,

$$d\tau_i(r,\psi) = \left[\frac{1}{2}\frac{\rho C_{l\alpha}cU_z(r,\psi)}{\varpi}(\theta_0 U_h^2(r,\psi) - U_z(r,\psi)U_h(r,\psi)) - \frac{1}{2}\frac{\rho C_{l\alpha}c(v_h^i + V_h)\sin\psi}{\varpi}(\theta_0 U_z(r,\psi)U_h(r,\psi) - U_z^2(r,\psi))\right]drd\psi,$$

$$d\tau_i(r,\psi) = \frac{U_z(r,\psi)}{\varpi}dT(r,\psi) - \frac{1}{2}\frac{\rho C_{l\alpha}c(v_h^i + V_h)\sin\psi}{\varpi}(\theta_0 U_z(r,\psi)U_h(r,\psi) - U_z^2(r,\psi))drd\psi. \tag{62}$$

From the equation for $dH_i$ (57), it can be shown that

$$\frac{1}{2}\rho C_{l\alpha} c(\theta_0 U_z(r,\psi) U_h(r,\psi) - U_z^2(r,\psi)) \sin\psi = dH_i - \frac{1}{2}\rho C_{l\alpha} c\left(\theta_0 U_h^2(r,\psi) - U_z(r,\psi) U_h(r,\psi)\right)\beta\cos\psi,$$
$$= dH_i(r,\psi) - dT\beta(\psi)\cos\psi.$$

Using the $d\tau_i(r,\psi)$ expression from (62)

$$d\tau_i(r,\psi) = \frac{1}{2}\rho C_{l\alpha} c(\theta_0 U_z(r,\psi) U_h(r,\psi) - U_z^2(r,\psi)) r dr d\psi,$$
$$= \frac{U_z(r\psi)}{\varpi} dT(r,\psi) - (dH_i(r,\psi) - dT(r,\psi)\beta(\psi)\cos\psi)\frac{(v_h^i + V_h)}{\varpi}.$$

Substituting $U_z(r,\psi) = v_z^i - V_z + r\dot\beta(\psi) + (v_h^i(r,\psi) + V_h(r,\psi))\beta(\psi)\cos\psi$, we obtain

$$d\tau_i(r,\psi) = \frac{v_z^i - V_z + r\dot\beta(\psi) + (v_h^i + V_h)\beta(\psi)\cos\psi}{\varpi} dT(r,\psi) - (dH_i - dT\beta\cos\psi)\frac{(v_h^i + V_h)}{\varpi},$$
$$= \frac{v_z^i - V_z + r\dot\beta(\psi)}{\varpi} dT(r,\psi) - dH_i \frac{(v_h^i + V_h)}{\varpi}.$$

The component as a result of flapping motion $\dot\beta(\psi)$

$$\frac{r}{\varpi} r\dot\beta(\psi) dT(r,\psi) = r dT(r,\psi)\frac{d\beta(\psi)}{d\psi}.$$

If there is no flapping hinge [6, pg. 101-102],

$$r dT(r,\psi)\frac{d\beta(\psi)}{d\psi} = 0.$$

Hence for fixed pitch quadrotor blades that do not have flapping hinges

$$\tau_i = (T\lambda - H_i\mu)R.$$

Substituting for $H_i$ and using

$$\tau = \tau_p + \tau_i,$$
$$= \tau_p + H_p\mu R + T\lambda R - H\mu R,$$

the torque is obtained as

$$\tau = \rho N_b c C_d R^4 \varpi^2 (1 + 3\mu^2)/8 + (T\lambda - H\mu)R. \tag{63}$$

The power can thus be easily obtained using $P = \tau\varpi$.

$$P = \frac{1}{8} N_b \rho c C_d R^4 \varpi^3 (1 + 3\mu^2) + (T\lambda - H\mu)\varpi R.$$

This power can be seen as consisting of profile power $(1 + 3\mu^2)$ term, power for axial motion $T\lambda$ and power for translational motion $H\mu$ of air relative to the rotor blade.

In summary, from blade element theory, we obtain the following equations for thrust $T$, power $P$ and in-plane $H$ (magnitude and direction) force for a constant chord and pitch rotor blade with a constant drag coefficient $C_d$ and zero-lift angle of attack with linear lift slope aerofoil are

$$T = \frac{1}{4} N_b \rho C_{l\alpha} c \varpi^2 R^3 \left[\frac{2}{3}\theta_0\left(1 + \frac{3}{2}\mu^2\right) - \lambda\right], \tag{64}$$

$$H = -\frac{1}{4}\rho C_{l\alpha} N_b c R^3 \varpi^2 \mu \left[\frac{C_d}{C_{l\alpha}} - \frac{1}{3}\theta_0\lambda + \lambda^2\right], \tag{65}$$

$$P = \frac{1}{8} N_b \rho c C_d R^4 \varpi^3 (1 + 3\mu^2) + (T\lambda - H\mu)\varpi R. \tag{66}$$

**Remark 9** *From the power model given by (66), the $P$ dependence on $H$ results in a $\mu^2$ term. For slow moving quadrotors with high disc loading, this term can be ignored.*

## 4.4 Lumped Models for Robotic Applications

In order to use these models for any practical robotic application, a lumped parameter model is required. By setting the following coefficients

$$c_0 = R,$$
$$c_1 = \frac{1}{4} N_b \rho C_{l_\alpha} c c_0^3,$$
$$c_2 = \frac{2}{3} \theta_0,$$

the thrust force is given by

$$T = c_1 \varpi^2 \left[ c_2 (1 + \frac{3}{2} \mu^2) - \lambda \right].$$

From the power equation in (66), if we set

$$c_3 = \frac{1}{8} \rho N_b c C_d c_0^4,$$

we get

$$P = c_3 \varpi^3 \left( 1 + 3\mu^2 \right) + (T\lambda - H\mu) \varpi c_0.$$

Therefore (65) can be rewritten as

$$H = -c_1 \mu \varpi^2 \left[ \frac{2 c_3}{c_1 c_0} - \frac{1}{3} \theta_0 \lambda + \lambda^2 \right].$$

So in summary, the lumped parameter model for thrust $T$, horizontal force $H$ and power $P$ and invoking Remark 8 for a rotor with

> **Assumptions**
> - Infinite aspect ratio with finite radius,
> - Constant chord and pitch,
> - Linear lift slope with zero-lift angle of attack aerofoil and
>
> **Forces and Power Relationships**
>
> $$T = c_1 \varpi^2 \left[ c_2 (1 + \frac{3}{2} \mu^2) - \lambda \right], \tag{67}$$
> $$H = -c_1 \mu \varpi^2 \left[ \frac{2 c_3}{c_1 c_0} - \frac{1}{3} \theta_0 \lambda + \lambda^2 \right], \tag{68}$$
> $$P = c_3 \varpi^3 \left( 1 + 3\mu^2 \right) + \left( T(\kappa \lambda^i - \lambda_z) - H(\kappa \mu^i + \mu_h) \right) \varpi c_0. \tag{69}$$

The coefficients $c_0, c_1, c_2, c_3$ can be determined by fitting the models to experimental data using linear regression or computed through measurements.

Though the assumptions of constant chord and pitch along with infinite aspect ratio and zero-lift angle of attack of the above models are generally not true for quadrotor rotors, they however represent a good starting point for modelling of quadrotor rotor blades. It should be noted that for helicopters, $\varpi$ is maintained at a constant value, changing power and the collective pitch setting using the swash plate mechanism results in changes in $T, H, \mu$ and $\lambda$. For quadrotors however, given the low mass moment of inertia blades indicates that fast dynamic responses of the rotor speed ($\varpi$) are achievable using an *electronic speed controller* (ESC). This results in changes in power and therefore $T$ and $\lambda$. $H$ and $\mu$ are generated through differential thrust changes in the four rotors that generate torques around $\vec{e}_1, \vec{e}_2$ and $\vec{e}_3$ of $\{B\}$.

In Section 5 and 6, we gradually remove some of the geometric (chord and pitch) and aerodynamic (lift and drag coefficients) assumptions made in this section to produce a more realistic model for quadrotor rotor blades.

# 5 Blade Element Theory for Constant Pitch, Constant Chord, Zero-lift Angle of Attack Aerofoil with Finite Aspect Ratio

Taking a keen look at quadrotor rotor blades, they are neither very long nor slender thus distinguishing them from helicopter rotor blades. Therefore the assumption of infinite aspect ratio (i.e. $AR = \infty$) or $K = 0$ cannot hold. Furthermore quadrotor rotors are not made out of thin flat sheets or aerofoils with no camber (or with zero-lift angle of attack) which may be true for helicopter rotors. The aerofoils on quadrotors have significant camber as they are required to produce higher thrust to torque ratios around hovering conditions. In this section, we take the blade element momentum theory modelling further by removing the zero-lift angle of attack aerofoil with infinite aspect ratio ($AR$) rotor assumption. The assumptions of constant chord and pitch are kept because they are true for certain quadrotors. An example of a quadrotor with such rectangular planform blades with cambered aerofoils is the Y4-Triangular configuration quadrotor [14].

## 5.1 The Thrust Force

Recall the elemental thrust from (50) and (51), hence

$$\mathrm{d}T(r,\psi) = \frac{1}{2}\rho U(r,\psi)^2 C_l(r,\psi)c(r)\mathrm{d}r\mathrm{d}\psi,$$
$$= \frac{1}{2}\rho U_h(r,\psi)^2(C_{l0} + C_{l\alpha}\alpha)c(r)\mathrm{d}r\mathrm{d}\psi. \tag{70}$$

The $C_{l\alpha}$ component is as found in Section 4.1. So we need only compute the $C_{l0}$ component of (70)

$$\mathrm{d}T_{Cl0}(r,\psi) = \frac{1}{2}\rho C_{l0} c(r) U_h(r,\psi)^2 \mathrm{d}r\mathrm{d}\psi.$$

Substituting for $U_h(r,\psi)$ and $c(r) = c$,

$$\mathrm{d}T_{Cl0}(r,\psi) = \frac{1}{2}\rho C_{l0} c \varpi^2 R^2 \left(\frac{r}{R} + \mu \sin\psi\right)^2 \mathrm{d}r\mathrm{d}\psi,$$

$$T_{Cl0} = \frac{N_b}{2\pi}\int_0^R \int_0^{2\pi} \frac{1}{2}\rho C_{l0} c \varpi^2 R^2 \left(\frac{r}{R} + \mu \sin\psi\right)^2 \mathrm{d}r\mathrm{d}\psi,$$

$$T_{Cl0} = \frac{1}{4}N_b \rho c C_{l0} R^3 \varpi^2 \left(\frac{2}{3} + \mu^2\right).$$

Adding this to the thrust due to $C_{l\alpha}$ found in Section 4.1,

$$T = \frac{1}{4}N_b \rho c R^3 \varpi^2 \left[C_{l\alpha}\left(\frac{2}{3}\theta_0\left(1 + \frac{3}{2}\mu^2\right) - \lambda\right) + C_{l0}\left(\frac{2}{3} + \mu^2\right)\right].$$

Comparing this to the thrust given by (64), there is an extra term $\propto C_{l0}\left(\frac{2}{3} + \mu^2\right)$ which clearly shows that any camber or lift offset for an angle of attack of zero will result in an increase in thrust. There is an additional thrust increment as a result of any translational motion.

## 5.2 The Horizontal H-Force

Recall from Section 4.2 where the $H$-force was shown to consist of both profile and lift induced components. If we consider first the profile contribution

$$H_\mathrm{P} = \frac{N_b}{2\pi}\int_0^{2\pi}\int_0^R \frac{1}{2}\rho U(r,\psi)^2 C_d(r,\psi) c(r) \sin\psi \mathrm{d}r\mathrm{d}\psi.$$

With $C_d(r,\psi) = C_{d0} + KC_l^2(r,\psi) = C_{d0} + K\left(C_{l0}^2 + 2C_{l0}C_{l\alpha}\alpha(r,\psi) + C_{l\alpha}^2\alpha^2(r,\psi)\right)$. The $C_{d0}$ and $KC_{l0}^2$ components are similar to that obtained in Section 4.2. The $2C_{l0}C_{l\alpha}\alpha(r,\psi)$ component is calculated as follows

$$\mathrm{d}H_{pcla}(r,\psi) = \rho K C_{l0} C_{l\alpha} c U_h^2(r,\psi) \left(\theta - \frac{U_z(r,\psi)}{U_h(r,\psi)}\right) \sin\psi \mathrm{d}r\mathrm{d}\psi,$$

$$= \rho K C_{l0} C_{l\alpha} c \left(\theta_0 U_h^2(r,\psi) - U_z(r,\psi)U_h(r,\psi)\right) \sin\psi \mathrm{d}r\mathrm{d}\psi,$$

$$= \rho K C_{l0} C_{l\alpha} c \varpi^2 R^2 \left(\theta_0 \left(\frac{r}{R} + \mu\sin\psi\right)^2 - \left(\frac{r}{R} + \mu\sin\psi\right)\left(\lambda + \frac{r}{R}\frac{\mathrm{d}\beta(\psi)}{\mathrm{d}\psi} + \mu\beta(\psi)\cos\psi\right)\right) \sin\psi \mathrm{d}r\mathrm{d}\psi,$$

$$H_{pcla} = \frac{1}{2}\rho N_b K C_{l0} C_{l\alpha} c \varpi^2 R^3 \left(\theta_0\mu - \lambda\mu - \frac{1}{3}a_1\right),$$

after setting $\mu^2 = 0$. For the $C_{l\alpha}^2 \alpha^2(r,\psi)$ component,

$$\mathrm{d}H_{pcla2}(r,\psi) = \frac{1}{2}\rho K C_{l\alpha}^2 \alpha^2 U_h^2(r,\psi) \sin\psi \mathrm{d}r\mathrm{d}\psi,$$

$$= \frac{1}{2}\rho K C_{l\alpha}^2 U_h^2(r,\psi) \left(\theta - \frac{U_z(r,\psi)}{U_h(r,\psi)}\right)^2 \sin\psi \mathrm{d}r\mathrm{d}\psi,$$

$$H_{pcla2} = \frac{N_b}{2\pi} \int_0^{2\pi} \int_0^R \frac{1}{2}\rho K C_{l\alpha}^2 R^2 \varpi^2 \left[\theta_0^2\left(\left(\frac{r}{R}\right)^2 + 2\left(\frac{r}{R}\right)\mu\sin\psi + \mu^2\sin^2\psi\right) - 2u_z\theta_0\left(\frac{r}{R} + \mu\sin\psi\right) + u_z^2\right]\sin\psi \mathrm{d}r\mathrm{d}\psi.$$

The integral of $u_z\theta_0 + \ldots + u_z^2$ is

$$= \left(-2u_z\theta_0\left(\frac{r}{R} + \mu\sin\psi\right) + u_z^2\right)\sin\psi \mathrm{d}r\mathrm{d}\psi,$$

$$= \left(-2\left(\lambda + \frac{r}{R}\frac{\mathrm{d}\beta(\psi)}{\mathrm{d}\psi} + \mu\beta(\psi)\cos\psi\right)\theta_0\left(\frac{r}{R} + \mu\sin\psi\right) + \left(\lambda + \frac{r}{R}\frac{\mathrm{d}\beta(\psi)}{\mathrm{d}\psi} + \mu\beta(\psi)\cos\psi\right)^2\right)\sin\psi \mathrm{d}r\mathrm{d}\psi,$$

$$= -2\mu\lambda\theta_0 - \frac{2}{3}a_1\theta_0 - 2\mu^2 a_0\theta_0 + \mu b_1\theta_0 + 2\lambda a_1 - b_1\lambda\mu.$$

Setting multiples of the flapping coefficients and $\mu^2$ to zero,

$$H_{cla2} = \frac{1}{4}N_b\rho K C_{l\alpha}^2 \varpi^2 R^3 \left(\frac{a_1\lambda}{2} - \frac{2a_1\theta_0}{5} + \lambda^2\mu - 2\lambda\mu\theta_0 + \mu\theta_0^2\right).$$

Adding all the components of $H_p$, we get

$$H_p = H_{pcd0} + H_{pcla} + H_{pcla2},$$

$$H_p = \frac{1}{4}N_b\rho c R^3 \varpi^2 \left(\mu C_{d0} + K C_{l0} C_{l\alpha}\left(\theta_0\mu - \lambda\mu - \frac{1}{3}a_1\right) + K C_{l\alpha}^2\left(\frac{a_1\lambda}{2} - \frac{2a_1\theta_0}{5} + \lambda^2\mu - 2\lambda\mu\theta_0 + \mu\theta_0^2\right)\right),$$

$$H_p = \frac{1}{4}N_b\rho c R^3 \varpi^2 \mu \left(C_{d0} + K C_{l0} C_{l\alpha}\left(\theta_0 - \lambda - \frac{2}{3}(4/3\theta_0 - \lambda)\right) + K C_{l\alpha}^2\left(\lambda(4/3\theta_0 - \lambda) - \frac{4\theta_0(4/3\theta_0 - \lambda)}{5} + \lambda^2 - 2\lambda\theta_0 + \theta_0^2\right)\right).$$

For the lift induced $H$-force, its $C_{l\alpha}$ component is the same as in Section 4.2 and the $C_{l0}$ component is obtained as follows

$$\mathrm{d}H_{icl0}(r,\psi) = \frac{1}{2}\rho c C_{l0}\left(\left(\frac{r}{R}\right)^2 + 2\left(\frac{r}{R}\right)\mu\sin\psi + \mu^2\sin^2\psi\right)\left(\beta(\psi)\cos\psi + \phi(r,\psi)\sin\psi\right)\mathrm{d}r\mathrm{d}\psi.$$

Consider first the $\phi(r,\psi)\sin\psi$ component which gives

$$H_{iclo\phi} = \frac{1}{4}N_b\rho c C_{l0}\varpi^2 R^3\left(\frac{a_1}{3} + \mu\lambda - \frac{a_1\mu^2}{4}\right).$$

Then consider the $\beta(\psi)$ component

$$dH_{icl0\beta}(r,\psi) = \frac{1}{2}\rho C_{l0}\left(\left(\frac{r}{R}\right)^2 + 2\left(\frac{r}{R}\right)\mu\sin\psi + \mu^2\sin^2\psi\right)(a_0 - a_1\cos\psi - b_1\sin\psi)\cos\psi\, drd\psi.$$

Taking the integral, it can be shown that

$$H_{icl0\beta} = -\frac{1}{4}\rho C_{l0}a_1\varpi^2 R^3\left(\frac{1}{3} + \frac{1}{4}\mu^2\right).$$

Summing all of these, we get

$$H_{i0} = H_{icl0\phi} + H_{icl0\beta},$$
$$H_{i0} = \frac{1}{4}\rho N_b c R^3 C_{l0}\varpi^2\left[\frac{a_1}{3} + \mu\lambda - \frac{a_1\mu^2}{4} - a_1\left(\frac{1}{3} + \frac{1}{4}\mu^2\right)\right],$$
$$H_{i0} = \frac{1}{4}\rho N_b c R^3 C_{l0}\varpi^2 \mu\lambda. \tag{71}$$

The total induced component is

$$H_i = H_{i0} + H_{i\mathbb{C}l\alpha}.$$
$$H_i = \frac{1}{4}N_b\rho C_{l\alpha}cR^3\varpi^2\mu\left[\lambda - \frac{1}{3}\theta_0\lambda + \lambda^2\right].$$

Hence the total $H$-force is given by

$$H = H_p + H_i$$
$$H = \frac{1}{4}N_b\rho C_{l\alpha}cR^3\varpi^2\mu\left[X + \lambda - \frac{1}{3}\theta_0\lambda + \lambda^2\right],$$

where

$$X = C_{d0} + KC_{l0}C_{l\alpha}\left(\theta_0 - \lambda - \frac{2}{3}(4/3\theta_0 - \lambda)\right) + KC_{l\alpha}^2\left(\lambda(4/3\theta_0 - \lambda) - \frac{4\theta_0(4/3\theta_0 - \lambda)}{5} + \lambda^2 - 2\lambda\theta_0 + \theta_0^2\right).$$

With camber or any nonzero lift at zero angle of attack, there is an additional term due to $C_{l0}$ (in the induced $H$-force) which causes an increment in the $H$-force. This further increases with translational motion. Given that $C_{l0}$ affects both $T$ and $H$ is the reason the blades are designed so that $\frac{C_l}{C_d}$ is maximum for the designed operating region. Furthermore the finite AR causes a further increase in this force. Hence to minimise the $H$-force, it is necessary to design blades with symmetric or $C_{l0} = 0$ aerofoils that are infinitely long and thin so as to have a very large AR and therefore low $K$ or $K \cong 0$. This is the principal reason helicopter rotors are designed to be long and slender. It should also be noted that very long blades may result in higher blade flapping $\beta(\psi)$ due to reduced rigidity and increased flexibility.

### 5.3 The Torque and Power

Recall from (60) in Section 4.3,

$$d\tau(r,\psi) = r\left(dD(r,\psi)\cos\phi(r,\psi) + dL(r,\psi)\cos\beta(\psi)\sin\phi(r,\psi)\right)drd\psi,$$
$$= \underbrace{rdD(r,\psi)}_{d\tau_P} + \underbrace{rdL(r,\psi)\phi(r,\psi)}_{d\tau_i(r,\psi)}drd\psi. \tag{72}$$

Considering again the profile torque contribution first,

$$\mathrm{d}\tau_p = rdD(r,\psi)\mathrm{d}r\mathrm{d}\psi,$$

$$= \frac{1}{2}\rho c C_d(r,\psi) U_h^2(r,\psi) r\mathrm{d}r\mathrm{d}\psi,$$

$$= \frac{1}{2}\rho c U_h^2(r,\psi)\left(C_{d0} + K\left(C_{l0}^2 + \underbrace{2C_{l0}C_{l\alpha}\alpha(r,\psi)}_{\tau_{pclocl}} + \underbrace{C_{l\alpha}^2\alpha^2(r,\psi)}_{\tau_{pcla2}}\right)\right)r\mathrm{d}r\mathrm{d}\psi,$$

The $C_{d0}$ and $KC_{l0}^2$ components are similar to the $C_d(r,\psi)$ obtained in Section 4.3. For the $2C_{l0}C_{l\alpha}\alpha(r,\psi)$ term,

$$\mathrm{d}\tau_{pclocl} = \left(\frac{1}{2}\rho c\right) 2KC_{l0}C_{l\alpha} U_h^2(r,\psi)\left(\theta_0 - \frac{U_z(r,\psi)}{U_h(r,\psi)}\right) r\mathrm{d}r\mathrm{d}\psi,$$

$$= \left(\frac{1}{2}\rho c\right) 2KC_{l0}C_{l\alpha}\left(\theta_0 U_h^2(r,\psi) - U_z(r,\psi)U_h(r,\psi)\right) r\mathrm{d}r\mathrm{d}\psi,$$

$$\tau_{pclocl} = \left(\frac{1}{4}N_b\rho c\right) 2KC_{l0}C_{l\alpha}\left(\theta_0\left(\frac{1}{2} + \frac{1}{2}\mu^2\right) - \frac{1}{3}\lambda\right),$$

$$\tau_{pclocl} = \left(\frac{1}{4}N_b\rho c\right) KC_{l0}C_{l\alpha}\left(\theta_0(1+\mu^2) - \frac{4}{3}\lambda\right).$$

It should be noted that in the computation of $\tau_{pclocl}$, the flapping terms occur as products (for e.g. $a_1 b_1$) which can be assumed zero as it has higher powers of $\mu$ and are thus considered negligible.

The $C_{l\alpha}^2$ component is computed by noting that the products of any two flapping coefficients result in higher powers of $\mu$ and are therefore negligible.

$$\mathrm{d}\tau_{pcla2} = \left(\frac{1}{2}\rho c\right) K U_h^2(r,\psi) C_{l\alpha}^2 \alpha^2 r\mathrm{d}r\mathrm{d}\psi,$$

$$= \left(\frac{1}{2}\rho c\right) K C_{l\alpha}^2 U_h^2\left(\theta_0 - \frac{U_z(r,\psi)}{U_h(r,\psi)}\right)^2 r\mathrm{d}r\mathrm{d}\psi,$$

$$= \left(\frac{1}{2}\rho c\right) K C_{l\alpha}^2 \left(\theta_0 U_h(r,\psi) - U_z(r,\psi)\right)^2 r\mathrm{d}r\mathrm{d}\psi,$$

$$= \left(\frac{1}{2}\rho c\right) K C_{l\alpha}^2 \varpi^2 R^2 \left(\theta_0^2 u_h^2(r,\psi) - 2u_z(r,\psi)u_h(r,\psi) + u_z^2(r,\psi)\right) r\mathrm{d}r\mathrm{d}\psi.$$

Hence

$$\tau_{pcla2} = \left(\frac{1}{8}N_b\rho c\right) K C_{l\alpha}^2 \varpi^2 R^4 \left(4\lambda^2 - 16/3\lambda\theta_0 - a_1\mu\lambda + 2\theta_0^2\right),$$

$$\tau_{pcla2} = \left(\frac{1}{8}N_b\rho c\right) K C_{l\alpha}^2 \varpi^2 R^4 \left(4\lambda^2 - 16/3\lambda\theta_0 + 2\theta_0^2\right).$$

If the profile torque due to $C_{d0}$ and $KC_{l0}^2$ is $\tau_{cd}$, the

$$\tau_p = \tau_{cd} + \tau_{pclocl} + \tau_{pcla2},$$

$$\tau_p = \frac{1}{8}\rho N_b c \varpi^2 R^4 \Bigg[\left(C_{d0} + KC_{l0}^2\right)(1+\mu^2) + 2KC_{l0}C_{l\alpha}\left(\theta_0(1+\mu^2) - \frac{4}{3}\lambda\right) +$$

$$KC_{l\alpha}^2\left(4\lambda^2 - 16/3\lambda\theta_0 + 2\theta_0^2\right)\Bigg].$$

In a similar manner to Section 4.3, it can be shown that $\tau_i = (T\lambda - H_i\mu)R$. Hence the total torque is

$$\begin{aligned}\tau &= \tau_p + \tau_i, \\ &= \tau_p + H_p\mu R + T\lambda R - H\mu R, \\ &= \frac{1}{8}\rho N_b c\varpi^2 R^4 Z + (T\lambda - H\mu)R,\end{aligned}$$

where

$$\tau_p = \frac{1}{8}\rho N_b c\varpi^2 R^4 \bigg[ \left(C_{d0} + KC_{l0}^2\right)\left(1 + \mu^2\right) + 2KC_{l0}C_{l\alpha}\left(\theta_0\left(1 + \mu^2\right) - \frac{4}{3}\lambda\right) + \\ KC_{l\alpha}^2\left(4\lambda^2 - 16/3\lambda\theta_0 + 2\theta_0^2\right) + 2\Delta_\tau\bigg],$$

and

$$\Delta_\tau = \mu^2\frac{1}{4}N_b\rho cR^3\varpi^2\mu\left(C_{d0} + KC_{l0}C_{l\alpha}\left(\theta_0 - \lambda - \frac{2}{3}(4/3\theta_0 - \lambda)\right) + KC_{l\alpha}^2\left(\lambda(4/3\theta_0 - \lambda) - \frac{4\theta_0(4/3\theta_0 - \lambda)}{5} + \lambda^2 - 2\lambda\theta_0 + \theta_0^2\right)\right).$$

Comparing this result to that obtained in Section 4.3, it can be deduced that an increase in camber results in a higher required torque which translates to increase required power. Furthermore, the finite AR of the rotor blades result in an increase in the $H$-force and power required.

So in summary based on Remark 8 for a rotor with

### Assumptions

- Nonzero lift angle of attack aerofoil with linear lift slope,
- Constant chord and pitch,
- Finite aspect ratio and

### Forces and Power Relationships

$$T = \frac{1}{4}N_b\rho cR^3\varpi^2\left[C_{l\alpha}\left(\frac{2}{3}\theta_0\left(1 + \frac{3}{2}\mu^2\right) - \lambda\right) + C_{l0}\left(\frac{2}{3} + \mu^2\right)\right], \tag{73}$$

$$H = -\frac{1}{4}N_b\rho C_{l\alpha}cR^3\varpi^2\mu\left[X + \lambda - \frac{1}{3}\theta_0\lambda + \lambda^2\right], \tag{74}$$

$$P = \frac{1}{8}\rho N_b c\varpi^3 R^4 Z + \left(T(\kappa\lambda^i - \lambda_z) - H(\kappa\mu^i + \mu_h)\right)\varpi R, \tag{75}$$

where

$$X = C_{d0} + KC_{l0}C_{l\alpha}\left(\theta_0 - \lambda - \frac{2}{3}(4/3\theta_0 - \lambda)\right) + KC_{l\alpha}^2\left(\lambda(4/3\theta_0 - \lambda) - \frac{4\theta_0(4/3\theta_0 - \lambda)}{5} + \lambda^2 - 2\lambda\theta_0 + \theta_0^2\right),$$

$$Z = \left(C_{d0} + KC_{l0}^2\right)\left(1 + 3\mu^2\right) + 2KC_{l0}C_{l\alpha}\left(2(\theta_0 - \lambda)\mu^2 + \theta_0(1 + \mu^2) - \frac{2}{3}\lambda\right) + \\ KC_{l\alpha}^2\left(2\theta_0(\theta_0 - 2\lambda)\mu^2 + \theta_0^2(1 + \mu^2) - \frac{8}{3}\lambda + 2\lambda^2\right) + \Delta_\tau,$$

and

$$\Delta_\tau = \mu^2\frac{1}{4}N_b\rho cR^3\varpi^2\mu\left(C_{d0} + KC_{l0}C_{l\alpha}\left(\theta_0 - \lambda - \frac{2}{3}(4/3\theta_0 - \lambda)\right) + KC_{l\alpha}^2\left(\lambda(4/3\theta_0 - \lambda) - \frac{4\theta_0(4/3\theta_0 - \lambda)}{5} + \lambda^2 - 2\lambda\theta_0 + \theta_0^2\right)\right).$$

If we set $K = 0$ or infinite $AR$ and $C_{l0} = 0$, (73) to (75) reduce to (64) to (66) respectively. It is now clear that high cambered (or any nonzero $C_{l0}$) aerofoils result in increased thrust with an increased $H$-force and required power. Any reduction in $K$ by increasing the $AR$ of the blades result in lower $H$-force and hence required power to maintain a desired $T$ for some advance ratio $\mu$. This is the reason helicopters have very long slender blades. Thus the aerofoil section geometry dictated by $C_{l0}$ and $AR$ of the blades must be carefully chosen to meet the design specification of the quadrotor if constant chord and pitch rotors are used.

# 6 Blade Element Momentum Theory for Ideal Rotor Blades

In this section, we develop further the models outlined in Section 4 and 5 for the majority of quadrotor blades for which the constant chord and pitch assumptions do not hold. These blades are designed to have ideal twist and chord in order to achieve optimality at hover and are otherwise referred to as ideal or optimum hovering rotor blades [8]. Similar to Section 5, the blades are considered to be made out of cambered aerofoils or have nonzero lift angle of attack sections with finite AR. Before going any further, it is worth defining an important parameter in the study of the dynamics of objects in fluids, the Reynolds number ($Re$).

**Definition 7** *The Reynolds number Re is a non-dimensional quantity defined as the ratio of inertial forces to viscous forces*
$$Re(r, \psi) = \frac{U_h(r, \psi) c(r)}{\nu},$$

where $\nu$ is the kinematic viscosity and is defined as the ratio of dynamic viscosity ($v_\mu$) to density ($\rho$) i.e. $\nu = v_\mu/\rho$. At room temperature, $v_\mu = 1.983 \times 10^{-5} Pa.s$, $\rho = 1.225 kg/m^3$ for air. Hence $\nu = 1.6188 \times 10^{-5} m^2/s$. If the tip velocity of the rotor is $52 m/s$ (Assumption 3.2) with tip chord $1 cm$, then the Reynolds number at the tip is

$$Re = \frac{52 \times 0.01}{10^{-5}} = 3.2123 \times 10^4.$$

This is the maximum Reynolds number for hover. Given that for the quadrotor under study, the maximum assumed velocity is $5 m/s$, implies that translational velocities will have little effect on this value. This value of Reynolds number is considered as low and the flow on the rotor is therefore laminar and is dominated by viscous forces [18, 19]. A second important parameter is the Mach number which is defined as

**Definition 8**
$$M = \frac{V}{V_{sound}}.$$

Assuming sea level operating conditions where the speed of sound $V_{sound} = 340.29 m/s$ implies that the tip Mach number is $M = 0.15$. This value of Mach number is said to be small and thus the assumption of incompressibility holds.

## 6.1 Ideal Rotors and Optimality

Recall from Section 3.6, that an optimal rotor is one with ideal chord and pitch defined by the following hyperbolic functions respectively
$$c(r) = \frac{c_{tip}}{r/R},$$
and
$$\theta(r) = \frac{\theta_{tip}}{r/R},$$

respectively. In which $C_{tip}, \theta_{tip}$ are the tip chord and pitch respectively. From [6, pg. 54] and [8, 15, 16], an optimal hovering rotor is a rotor that is capable of maintaining

1. Constant spanwise Reynolds number,

2. Constant spanwise angle of attack,

3. Constant induced velocity to reduce induce power

at hover. Given that $\theta \to 90°$ is possible at the hub where $r \to 0$ for the hyperbolic geometry, implies that the angle of attack $\alpha \to 90°$ as $r \to 0$. Hence the linear coefficient of lift relationship $C_l = C_{l0} + C_{l\alpha}\alpha$ which if $\alpha_s$ is the stall angle holds only when $|\alpha| \leq \alpha_s$. To prove optimality, consider the rotor at hover for which $|V| = 0$. The Reynolds number at a blade section is given by

$$Re(r) = \frac{\varpi r \frac{c_{tip}}{r}}{\nu} = \frac{\varpi c_{tip}}{\nu},$$

which is constant. Hence the first condition of optimality at hover is satisfied by choosing an ideal chord. The angle of attack for small angles based on Assumption 3.3 defined previously is given by

$$\alpha(r,\psi) = \theta(r) - \frac{U_z(r,\psi)}{U_h(r,\psi)}.$$

At hover, $\mu = 0$. Hence $U_z(r,\psi) = \lambda \varpi R$. To obtain a constant spanwise angle of attack, ideal twist results in

$$\alpha(r) = \theta_{tip}\frac{R}{r} - \lambda\frac{R}{r},$$
$$= (\theta_{tip} - \lambda)\frac{R}{r}.$$

Hence at every section along the span, the angle of attack $\alpha$ is a constant which we denote by $\alpha_r$ and therefore satisfies the second condition of optimality. This angle is chosen to maximise $\frac{C_l}{C_d}$ [6]. Given that we choose $\alpha = \alpha_r$, the pitch at every section is given by

$$\theta(r) = \alpha_r + \frac{v_z^i}{\varpi r},$$

The most optimum induced velocity for the reduction of induced power is a constant value. This was pointed out in Remark 7. From Section 5, the effect of $C_{l0}$ is to cause an offset in $T$ thereby increasing or decreasing it when $\mu = 0$. By ignoring this for simplicity of analysis, the elemental thrust is given by

$$dT = \frac{1}{2}\rho\varpi^2 r^2 C_{l\alpha}\alpha c(r)dr,$$
$$= \frac{1}{2}\rho\varpi^2 r^2 C_{l\alpha}(\theta_{tip} - \lambda)\frac{R}{r}c(r)dr, \tag{76}$$

and from momentum theory (Section 2.3),
$$dT = 4\pi\rho(v_z^i)^2 dr \tag{77}$$

Equating (76) to (77) for a constant induced velocity, one requires $c(r) = \frac{c_{tip}}{r/R}$. Therefore condition 3 is satisfied by choosing an ideal chord. The lift and therefore thrust is thus given by

$$dT(r,\psi) \cong dL(r,\psi) = \frac{1}{2}\rho U(r,\psi)^2 C_l(r,\psi)c(r)drd\psi,$$
$$= \frac{1}{2}\rho c_{tip}\varpi^2 R(C_{l0} + C_{l\alpha}\alpha_r)rdrd\psi,$$
$$T = \frac{1}{4}N_b\rho c_{tip}\varpi^2 R^3 (C_{l0} + C_{l\alpha}\alpha_r).$$

The H force is zero as $\mu = 0$. The profile torque is given by

$$\mathrm{d}\tau_p(r,\psi) = \frac{1}{2}\rho U(r,\psi)^2 C_d(r,\psi) c(r) r \mathrm{d}r\mathrm{d}\psi,$$

$$\mathrm{d}\tau_p(r,\psi) = \frac{1}{2}\rho c_{tip} R\varpi^2 r^2 \left(C_{d0} + KC_l^2\right)\mathrm{d}r\mathrm{d}\psi,$$

$$\tau_p = \frac{1}{6}N_b\rho c_{tip}R^4\varpi^2 \left(C_{d0} + KC_l^2\right).$$

Given that $C_l(r,\psi)$ is a constant as $\alpha$ is also constant, the lift induced torque is thus

$$\mathrm{d}\tau_i(r,\psi) = dL(r,\psi)\phi(r,\psi) r \mathrm{d}r\mathrm{d}\psi,$$

$$\mathrm{d}\tau_i(r,\psi) = \frac{1}{2}\rho c_{tip} U_h^2 C_l \phi R \mathrm{d}r\mathrm{d}\psi,$$

$$\mathrm{d}\tau_i = \frac{1}{2}\rho c_{tip} R^2 \varpi^2 r \left(C_{l0} + C_{l\alpha}\alpha_r\right)\lambda \mathrm{d}r\mathrm{d}\psi,$$

$$\tau_i = \frac{1}{4}N_b\rho c_{tip} R^4 \varpi^2 \left(C_{l0} + C_{l\alpha}\alpha_r\right)\lambda.$$

Comparing this to the thrust equation and in a similar manner to Section 4.3,

$$\tau_i = T\lambda R.$$

Hence for the optimum hovering rotor at hover,

$$T = \frac{1}{4}N_b\rho\varpi^2 R^3 c_{tip}\left(C_{l0} + C_{l\alpha}\alpha_r\right), \tag{78}$$

$$H = 0, \tag{79}$$

$$P = \frac{1}{6}N_b\rho c_{tip} R^4 \varpi^3 \left(C_{d0} + K\left(C_{l0} + C_{l\alpha}\alpha_r\right)^2\right) + T\lambda\varpi R. \tag{80}$$

Therefore maximum thrust is generated for a very low power. This is seen through the choice of $\alpha_r$ such that

$$\frac{C_{l0} + C_{l\alpha}\alpha_r}{C_{d0} + K\left(C_{l0} + C_{l\alpha}\alpha_r\right)^2},$$

is maximum. In the design of this optimum rotor, $\alpha_r$ is chosen based on the aerofoil used in making the blades and then the tip chord ($c_{tip}$) and pitch ($\theta_{tip}$) based on the radius ($R$) of the rotor. In reality, the rotor is required to not only hover or perform axial motion but also translational motion. The next subsection looks at the development of models for $T, H$ and $\tau$ (hence $P$) for a rotor with some geometric variation from ideal to enhance general quadrotor flight. These blades we refer to as "near ideal" rotor blades.

## 6.2 Generalised Blade Element Theory for "Near Ideal" Rotor Blades

In this subsection, the models for $T, H$ and $\tau$ (hence power) for "near ideal" rotors operating in non-hovering conditions are developed. We define "near ideal" rotors as rotors with ideal twist and chord up to a point from where the pitch flattens and the chord is curved inwards. By considering manufacturing of the ideal blades imposed by physical constraints on chord and pitch and the considerable $H$ force generated when $|V_h| \neq 0$, we will propose for there to be some geometric variation on pitch and chord from the hyperbolic geometry of the rotor around the hub. This leads to a trade-off between optimal thrust generation and the $H$-force. This trade-off is a design requirement based on the desired flight envelope.

Taking another look at the Reynolds number at a blade section $r$ and azimuth $\psi$

$$Re(r,\psi) = \frac{(\varpi r + (V_h + v_h^i)\sin\psi)c_{tip}R/r}{\nu},$$

$$= \frac{\varpi R\left(1 + \frac{R}{r}\mu\sin\psi\right)c_{tip}}{\nu}.$$

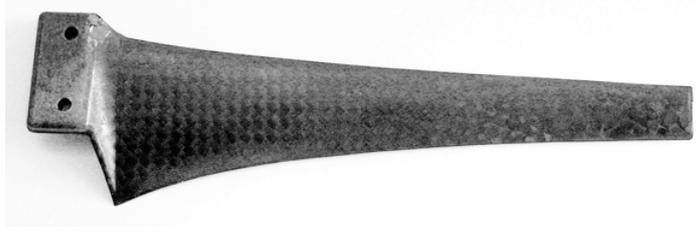

Figure 12: A near ideal rotor used on the ANU X4-Flyer. Rotor with optimum twist and chord with non-ideal geometry close to the hub [15, 16].

Any nonzero value of $\mu$ results in a very large $Re$ as $r \to 0$ near the hub. The angle of attack of the blade section then becomes

$$\alpha(r,\psi) = \theta(r) - \frac{U_z(r,\psi)}{U_h(r,\psi)}.$$

Hence at every section, $\alpha$ is no longer constant and therefore there is no guarantee that $|\alpha(r,\psi)| \leq \alpha_s$. Therefore the linear relationship between $C_l(r,\psi)$ and $\alpha(r,\psi)$ given by (37) is not guaranteed to hold. For a practical rotor, we know that $\theta(r)$, $c(r)$ have meaningful dimensional limits and therefore the Reynolds number $Re$ has an upper bound. We remodel the angle of attack by

$$\alpha(r,\psi) = \tan^{-1}\left(\frac{\theta_{tip}}{r/R} - \frac{u_z(r,\psi)}{u_h(r,\psi)}\right).$$

This is the original $\alpha(r,\psi)$ model as at low angles, $\tan \alpha(r,\psi) = \alpha(r,\psi)$. An example of a "near ideal" rotor blade is that shown in Figure 12 developed for and used on the ANU X4-Flyer [15, 16].

We use the $C_l$ model for high angles of attack proposed by Pucci [18, 19]. This is validated by the fact that the obtained $Re = 3.2 \times 10^4$ and Mach number $M = 0.15$ ensure that the flow is both laminar and incompressible and satisfies the conditions $Re < 160 \times 10^4$ and $M < 0.3$. The models for $C_l(r,\psi)$ and $C_d(r,\psi)$ in relation to $\alpha(r,\psi)$ are given by [18, 19]

$$C_l(r,\psi) = C_{l0} + C_{l\alpha}\alpha(r,\psi), \quad |\alpha(r,\psi)| \leq \alpha_s, \tag{81}$$

$$C_l(r,\psi) = C_2 \sin 2\alpha(r,\psi), \quad |\alpha(r,\psi)| > \alpha_s, \tag{82}$$

$$C_d(r,\psi) = C_{d0} + KC_l^2(r,\psi). \tag{83}$$

where $C_2$ is some positive constant. This model is based on experimental results from which it has been observed that at angles of attack beyond the stall angle $\alpha_s$, every aerofoil behaves like a flat plate with the boundary layer thickness increasing and the flow is detached from the rotor. From (81),(82) and (83), the elemental thrust therefore becomes

$$dT(r,\psi) = \frac{1}{2}\rho U_h^2(r,\psi) c_{tip} \left(C_2 \sin 2\alpha\right) \frac{R}{r} dr d\psi,$$

$$dT(r,\psi) = \frac{1}{2}\rho c_{tip} \frac{R}{r} (\varpi r + (V_h + v_h^i) \sin \psi)^2 C_2 \sin\left[2\tan^{-1}\left(\frac{\theta_{tip}}{r/R} - \frac{u_z(r,\psi)}{u_h(r,\psi)}\right)\right] dr d\psi,$$

$$dT(r,\psi) = \frac{1}{2}\rho c_{tip} \varpi^2 R^3 \left(\frac{r}{R^2} + 2\frac{1}{R}\mu \sin \psi + \frac{1}{r}\mu^2 \sin^2 \psi\right) C_2 \sin\left[2\tan^{-1}\left(\frac{\theta_{tip}}{r/R} - \frac{u_z(r,\psi)}{u_h(r,\psi)}\right)\right] dr d\psi. \tag{84}$$

Irrespective of the lift coefficient ((81) or (82)) model used, the thrust force becomes very large for any nonzero value of $\mu$. However, with the possibility of very high angles of attack around the hub, it is likely for the thrust to become zero depending on the ratio at which the rate of $r \to 0$ and $\alpha(r,\psi) \to 90°$.

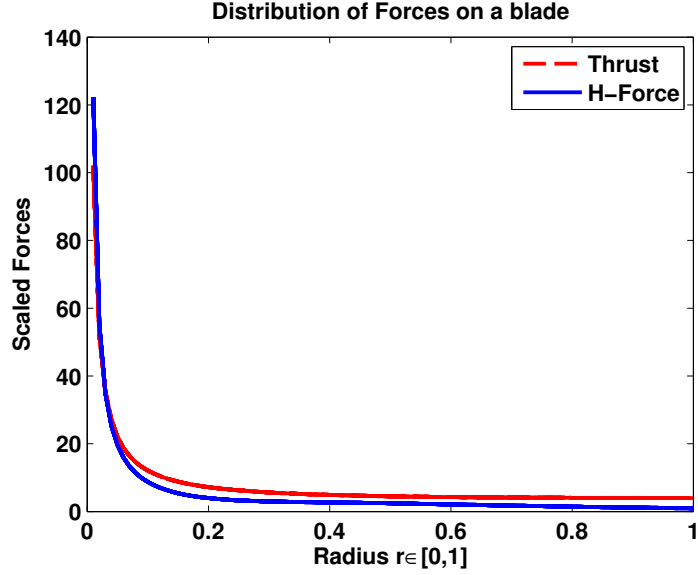

Figure 13: Scaled thrust and $H$-force spanwise distribution for ideal rotors. For more than 70% of the outer section of the blade, these forces can be approximated by constant forces.

In the determination of the $H$-force, first consider the profile drag contribution

$$\begin{aligned} dH_p(r,\psi) &= \frac{1}{2}\rho U_h(r,\psi)^2 C_d(r,\psi) c(r) \sin\psi \, dr \, d\psi, \\ &= \frac{1}{2}\rho c_{tip} \frac{R}{r} U_h(r,\psi)^2 \left(C_{d0} + K C_l^2(r,\psi)\right) \sin\psi \, dr \, d\psi. \end{aligned} \tag{85}$$

As $r \to 0$, implicit with (85), $dH_p(r,\psi)$ becomes very large or $dH_p(r,\psi) \to 0$ if $\mu = 0$ similar to the hover analysis. Consider now the torque which was shown before to consist of profile and torque as a result of the generation of $T$ and $H$. With $H$ becoming very large implies that the torque and hence power requirement for translation becomes too high. This also confirms optimality for maximum thrust generation (theoretically unbounded) at hover when $\mu = 0$ as the required power is upper bounded with $H = 0$. Hence the above set of equations reduce to that for the optimum hovering rotor at hover. If we set all the constants to 1 including $\sin\psi$ and ignoring the small flapping terms, Figure 13 shows $T$ and $H$ distributions across the span of an ideal rotor.

From Figure 13, it becomes obvious that a trade-off has to be made between $H$ and $T$ to produce a "near ideal/optimum" quadrotor rotor. The trade-off leads to a slight variation in geometry from the optimal rotor; one that is easily manufactured with significant reduction in $H$. Figure 14 shows an ideal rotor chord distribution along with a possible trade-off geometry which is similar to that shown in Figure 12 that was designed for the X4-Flyer [15, 16]. In addition, many other quadrotors have variations of these "near ideal" blades. Century Neo 860C is one such example [7].

Figure 13 shows that for more than 70% of the rotor, $T$ and $H$ (hence $\tau$) are almost constant. By modifying the section of the rotor around the hub, it is possible to generate a total thrust and horizontal force that will have an effective total as that of a constant thrust or H-force extending to the hub. The $T/H$ ratio from (84) and (85) for this region is a function of the mechanical connection to the hub, the mechanical properties such as aeroelasticity of the blade material and a specified value that is a performance criteria. This specified ratio depends on whether the quadrotor is designed for heavy lifting and near hovering flights or high speed and fast manoeuvring flights. In addition, it also depends on whether the vehicle will be flying in confined spaces such as near walls that can result in difficulty in control due to the significant changes in the generated $H$-force. From this discussion, we can make the following assumptions.

**Assumption 6.1** *We assume that the elemental forces and torque $(dT(r,\psi), dH(r,\psi), d\tau(r,\psi))$ have a constant value across the outer 70% region of the rotor. We also assume that the blade is designed such that for the*

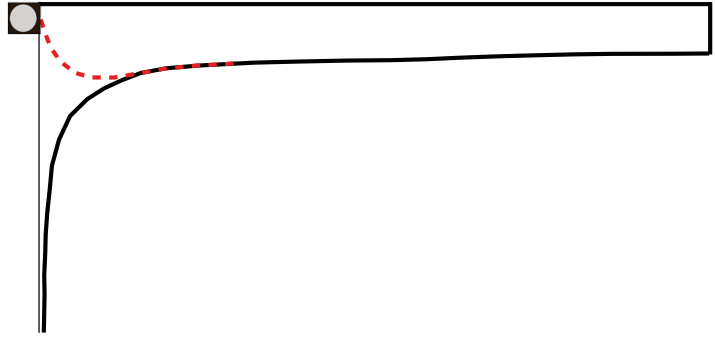

Figure 14: Geometry of an ideal (red) and a possible "near ideal" chord (black) length distribution.

*remaining 30% around the hub, the net effect can be modelled by the same constant elemental* $dT(r,\psi), dH(r,\psi)$ *and* $d\tau(r,\psi)$ *extending to the hub at* $r = 0$.

The validity of the assumption is the fact that $dT(r,\psi), dH(r,\psi)$ and therefore $d\tau(r,\psi)$ are constant for more than 70% of the rotor as shown in Figure 13. A consequence of Assumption 6.1 is that $|\alpha(r,\psi)| \leq \alpha_s$ and therefore the linear lift coefficient relationship (37) can be used. With this assumption, models for $T, H$ and $\tau$ are developed for the "near ideal" quadrotor blade in the next subsections. These models are based on equations presented in Section 4 and 5

### 6.2.1 The Thrust $T$ Model for "Near Ideal" Rotors

From the $dT(r,\psi)$ model presented in Section 4, the thrust is given by

$$
\begin{aligned}
dT(r,\psi) &= \frac{1}{2}\rho U_h^2(r,\psi) C_l(r,\psi) c_{tip} \frac{R}{r} drd\psi, \\
&= \frac{1}{2}\rho c_{tip}\left(\varpi r + (V_h + v_h^i)\sin\psi\right)^2 \left(C_{l0} + C_{l\alpha}\alpha(r,\psi)\right) \frac{R}{r} drd\psi, \\
&= \frac{1}{2}\rho c_{tip}\varpi^2 R^2 \left(\frac{r}{R} + \mu\sin\psi\right)^2 \left(C_{l0} + C_{l\alpha}\left(\theta_{tip}\frac{R}{r} - \frac{u_z}{r/R + \mu\sin\psi}\right)\right)\frac{R}{r}drd\psi, \\
&= \frac{1}{2}\rho c_{tip}\varpi^2 R^2 \\
& \quad (1+\mu\sin\psi)^2\left(C_{l0} + C_{l\alpha}\left(\theta_{tip} - \frac{\lambda + (a_1\sin\psi - b_1\cos\psi) + \mu(a_0 - a_1\cos\psi - b_1\sin\psi)\cos\psi}{1+\mu\sin\psi}\right)\right)drd\psi, r = R, \\
T &= \frac{1}{4}N_b\rho c_{tip}R^3\varpi^2\left(C_{l0}\left[2+\mu^2\right] + C_{l\alpha}\left[\theta_{tip}(2+\mu^2) - 2\lambda\right]\right).
\end{aligned}
$$

### 6.2.2 H-Force for "Near Ideal" Rotors

We start our modelling by looking at the profile contribution which was defined by (54) and is given by

$$
\begin{aligned}
dH_p(r,\psi) &= \frac{1}{2}\rho U_h^2(r,\psi) C_d(r,\psi) c(r)\sin\psi drd\psi, \\
&= \frac{1}{2}\rho c_{tip} U_h^2(r,\psi)\left((C_{d0} + KC_{l0}^2) + K\left(2C_{l0}C_{l\alpha}\alpha(r,\psi) + C_{l\alpha}^2\alpha^2(r,\psi)\right)\right)\frac{R}{r}\sin\psi drd\psi, \\
&= \frac{1}{2}\rho c_{tip}\varpi^2 R^2(1+\mu\sin\psi)^2\left((C_{d0} + KC_{l0}^2) + K\left(2C_{l0}C_{l\alpha}\alpha(r,\psi) + C_{l\alpha}^2\alpha^2(r,\psi)\right)\right)\sin\psi d\psi, \ r = R.
\end{aligned}
$$

Hence,

$$
\begin{aligned}
H_p = \frac{1}{2}N_b\rho c_{tip}\varpi^2 R^3\mu\big(&(C_{d0} + KC_{l0}^2) + KC_{l0}C_{l\alpha}(8\mu\theta_{tip} - 4a_1 - 4\lambda\mu) + \\
&KC_{l\alpha}^2\left(4\lambda a_1 - 4a_1\theta_{tip} + 4\mu\theta_{tip}^2 - 4\mu\theta_{tip}\right)\big).
\end{aligned}
$$

For the lift induced component

$$dH_i(r,\psi) = \frac{1}{2}\rho U_h^2(r,\psi)C_l(r,\psi)c(r)\left(\beta(\psi)\cos\psi + \phi(r,\psi)\sin\psi\right)drd\psi,$$
$$= \frac{1}{2}\rho U_h^2(r,\psi)c_{tip}\left(C_{l0} + C_{l\alpha}\alpha(r,\psi)\right)\left(\beta(\psi)\cos\psi + \phi(r,\psi)\sin\psi\right)drd\psi, \ r = R.$$

Hence,

$$H_i = \frac{1}{4}N_b\rho c_{tip}\varpi^2 R^3\left[C_{l0}\mu\lambda + C_{l\alpha}\left(\theta_{tip}\mu\lambda - a_1\lambda\right)\right],$$
$$H_i = \frac{1}{4}N_b\rho c_{tip}\varpi^2 R^3\mu\left[C_{l0}\lambda + C_{l\alpha}\left(\theta_{tip}\lambda - 2\lambda(\frac{4\theta_{tip}}{3} - \lambda)\right)\right].$$

Therefore the $H$-force is given by

$$H = \frac{1}{2}N_b\rho c_{tip}\varpi^2 R^3\mu\left[(C_{d0} + KC_{l0}^2) + KC_{l0}C_{l\alpha}(8\mu\theta_{tip} - 4a_1 - 4\lambda\mu) + KC_{l\alpha}^2\left(4\lambda a_1 - 4a_1\theta_{tip} + 4\mu\theta_{tip}^2 - 4\mu\theta_{tip}\right) + \frac{1}{2}X\right],$$

where

$$X = C_{l0}\lambda + C_{l\alpha}\left(\theta_{tip}\lambda - 2\lambda(\frac{4\theta_0}{3} - \lambda)\right).$$

### 6.2.3 Torque ($\tau$) and Power ($P$) for "Near Ideal" Rotors

The profile torque defined earlier is given by

$$d\tau_p(r,\psi) = \frac{1}{2}\rho U_h^2(r,\psi)\left(\underbrace{C_{d0}}_{\tau_{pcd0}} + K\left(C_{l0}^2 + \underbrace{2C_{l0}C_{l\alpha}\alpha(r,\psi)}_{\tau_{plocl}} + \underbrace{C_{l\alpha}^2\alpha^2(r,\psi)}_{\tau_{pcla2}}\right)\right)c_{tip}\frac{R}{r}rdrd\psi.$$

The $C_{d0}$ component

$$d\tau_{pcd0}(r,\psi) = \frac{1}{2}\rho U_h^2(r,\psi)C_{d0}c_{tip}Rdrd\psi,$$
$$\tau_{pcd0} = \frac{1}{4}N_b\rho c_{tip}C_{d0}R^4\varpi^2\left(2 + \mu^2\right),$$

which is similar to the $KC_{l0}^2$ component. For the $KC_{l0}C_{l\alpha}$ component,

$$d\tau_{pclocla}(r,\psi) = \rho c_{tip}KC_{l0}C_{l\alpha}U_h^2(r,\psi)\left(\theta_{tip} - \frac{U_z(r,\psi)}{U_h(r,\psi)}\right)Rdrd\psi,$$
$$d\tau_{pclocla}(r,\psi) = \rho c_{tip}KC_{l0}C_{l\alpha}R\left(\theta_{tip}U_h^2(r,\psi) - U_z(r,\psi)U_h(r,\psi)\right)drd\psi,$$
$$\tau_{pclocla} = \frac{1}{2}\rho N_b c_{tip}KC_{l0}C_{l\alpha}\varpi^2 R^4\left(\theta_{tip}\left(2 + \mu^2\right) - 2\lambda\right).$$

In the preceding calculations the fact that $\mu a_0, \mu a_1, \mu^2 b_1$ are negligible was used. For the $C_{l\alpha}^2$ component,

$$d\tau_{pcla2}(r,\psi) = \frac{1}{2}\rho K U_h^2(r,\psi)C_{l\alpha}^2\alpha^2 c(r)rdrd\psi,$$
$$d\tau_{pcla2}(r,\psi) = \frac{1}{2}\rho K c_{tip}RC_{l\alpha}^2 U_h^2(r,\psi)\left(\theta_{tip} - \frac{U_z(r,\psi)}{U_h(r,\psi)}\right)^2 drd\psi,$$
$$d\tau_{pcla2}(r,\psi) = \frac{1}{2}\rho K c_{tip}RC_{l\alpha}^2\left(\theta_{tip}U_h(r,\psi) - U_z(r,\psi)\right)^2 drd\psi,$$
$$d\tau_{pcla2}(\psi) = \frac{1}{2}\rho K c_{tip}C_{l\alpha}^2\varpi^2 R^3\left(\theta_{tip}^2(1 + \mu\sin\psi)^2 - 2\theta_{tip}u_z(r,\psi)(1 + \mu\sin\psi) + u_z(r,\psi)^2\right)drd\psi, r = R,$$
$$\tau_{cla2} = \frac{1}{2}\rho K N_b c_{tip}RC_{l\alpha}^2\varpi^2 R^4\left(2\theta_{tip}^2\left(1 + \mu^2\right) - 4\theta_{tip}\lambda + 2\lambda^2\right).$$

Hence the total profile torque is given by

$$\tau_p = \tau_{pcd0} + \tau_{pclocla} + \tau_{pcla2},$$

$$\tau_p = \frac{1}{4} N_b \rho c_{tip} R^4 \varpi^2 \left( (C_{d0} + KC_{l0}^2)(2+\mu^2) + 2KC_{l0}C_{l\alpha}\left(\theta_{tip}(2+\mu^2) - 2\lambda\right) + 2KC_{l\alpha}^2\left(2\theta_{tip}^2(1+\mu^2) - 4\theta_{tip}\lambda + 2\lambda^2\right) \right).$$

Similar to Section 4.3, it can be shown that $\tau_i = (T\lambda - H_i\mu)R$. Hence,

$$\tau = \frac{1}{8} N_b \rho c_{tip} \varpi^2 R^4 Z + (T\lambda - H\mu)R,$$

where

$$Z = 2\bigg( (C_{d0} + KC_{l0}^2)(2+3\mu^2) + 2KC_{l0}C_{l\alpha}\left((2\theta_{tip} - \lambda)\mu^2 + \left(\theta_{tip}(2+\mu^2) - 2\lambda\right)\right)$$
$$+ 2KC_{l\alpha}^2\left(2\theta_{tip}(\theta_{tip}-\lambda)\mu^2 + 2\theta_{tip}^2(1+\mu^2) - 4\theta_{tip}\lambda + 2\lambda^2\right) \bigg) + 2\Delta_\tau,$$

and

$$\Delta_\tau = \mu^2\left((C_{d0} + KC_{l0}^2) + KC_{l0}C_{l\alpha}(8\mu\theta_{tip} - 4a_1 - 4\lambda\mu) + KC_{l\alpha}^2\left(4\lambda a_1 - 4a_1\theta_{tip} + 4\mu\theta_{tip}^2 - 4\mu\theta_{tip}\right)\right).$$

In summary the models for $T, H$ and $P$ for a "near-ideal" rotor and from Remark 8 with

### Assumptions

- Nonzero $C_{l0}$ such as cambered NACA aerofoils,
- "Near ideal/optimum" chord and twist to a point near the hub,
- Finite aspect ratio and

### Forces and Power Relationships

$$T = \frac{1}{4} N_b \rho c_{tip} R^3 \varpi^2 \left( C_{l0}[2+\mu^2] + C_{l\alpha}[\theta_{tip}(2+\mu^2) - 2\lambda] \right), \qquad (86)$$

$$H = -\frac{1}{2} N_b \rho c_{tip} \varpi^2 R^3 \mu \left[ \zeta + \frac{1}{2} X \right], \qquad (87)$$

$$P = \frac{1}{8} \rho N_b c \varpi^3 R^4 Z + \left( T(\kappa\lambda^i - \lambda_z) - H(\kappa\mu^i + \mu_h) \right) \varpi R, \qquad (88)$$

where

$$\zeta = \left[ (C_{d0} + KC_{l0}^2) + KC_{l0}C_{l\alpha}(8\mu\theta_{tip} - 4a_1 - 4\lambda\mu) + KC_{l\alpha}^2\left(4\lambda a_1 - 4a_1\theta_{tip} + 4\mu\theta_{tip}^2 - 4\mu\theta_{tip}\right) \right],$$

$$X = \left[ C_{l0}\lambda + C_{l\alpha}\left(\theta_{tip}\lambda - 2\lambda(\frac{4\theta_0}{3} - \lambda)\right) \right],$$

$$Z = 2\bigg( (C_{d0} + KC_{l0}^2)(2+3\mu^2) + 2KC_{l0}C_{l\alpha}\left((2\theta_{tip} - \lambda)\mu^2 + \left(\theta_{tip}(2+\mu^2) - 2\lambda\right)\right)$$
$$+ 2KC_{l\alpha}^2\left(2\theta_{tip}(\theta_{tip}-\lambda)\mu^2 + 2\theta_{tip}^2(1+\mu^2) - 4\theta_{tip}\lambda + 2\lambda^2\right) \bigg) + 2\Delta_\tau,$$

$$\Delta_\tau = \mu^2\left((C_{d0} + KC_{l0}^2) + KC_{l0}C_{l\alpha}(8\mu\theta_{tip} - 4a_1 - 4\lambda\mu) + KC_{l\alpha}^2\left(4\lambda a_1 - 4a_1\theta_{tip} + 4\mu\theta_{tip}^2 - 4\mu\theta_{tip}\right)\right).$$

If we design a rotor such that $c_{tip} = c$, $\theta_{tip} = \theta_{tip}$, then the "near ideal" rotor produces far more $T, H$ and as such requires more $P$ due to the added twisted material than the rotors studied in Section 4 and 5.

# 7 Conclusion

In this report, we have developed models for the aerodynamic forces (thrust and H-force) and torque and hence power using momentum and blade element theories for quadrotor rotor blades from a robotics perspective. By choosing to model in the body fixed frame, we proposed and model the horizontal or H-force using momentum theory. By looking at the different vortex states in axial flights and the inability of momentum theory to account for the non-constant elemental velocities and forces lead to the use of blade element theory. Using blade element methods, we developed models for different blade geometries and aerodynamic properties of the aerofoils of rotor blades used on quadrotors. Finally, in the case of the ideal hovering rotor with hyperbolic geometry used on the majority of quadrotors, we showed that these rotors are only practical with geometric modifications that can make them meet the desired flight envelope and enhance manufacturing of the blades.

# A  Origin of the Constant Thrust and Torque Model

No theoretical development of the aerodynamics of quadrotor rotor blades is complete without a discussion on the currently used explicit constant thrust and torque to rotor speed free air models. These models have been used in performing some of the most impressive quadrotor manoeuvres which include flying through hoops [12], multiple flips [10] and ball catching [20]. From momentum theory, the static relationships between thrust, torque and power to rotor speed $\varpi$ can be derived. The so-called static models are based on the hover condition assumption. Starting with the momentum theory model for thrust and power using (11) and (13) and setting the velocity of the vehicle $V = 0$,

$$T = 2\rho A (v_z^i)^2,$$
$$P = 2\rho A (v_z^i)^3.$$

From Definition 2, $v_z^i = \lambda^i \varpi R$, substituting into the above,

$$T = 2\rho A R^2 (\lambda^i)^2 \varpi^2,$$
$$P = 2\rho A R^3 (\lambda^i)^3 \varpi^3.$$

Setting $C_T = 2\rho A R^2 (\lambda^i)^2$, $C_P = 2\rho A R^3 (\lambda^i)^3$, the static free air model for thrust, torque and therefore power are

$$T = C_T \varpi^2, \tag{89}$$
$$\tau = C_Q \varpi^2, \tag{90}$$
$$P = C_P \varpi^3. \tag{91}$$

Noting that $P = \tau \varpi$, thus $C_Q = C_P$. From experimental results, the thrust equation (89) has been modified to fit experimental data. The modified equation is [5]

$$T = C_{T0} \varpi + C_{T1} \varpi^2,$$

where $C_{T0}$ and $C_{T1}$ are constants.